\documentclass[pre,letterpaper,floatfix]{revtex4-1}
\pdfoutput=1
\usepackage{graphicx}
\usepackage{epstopdf}
\usepackage{amsmath}
\usepackage{amssymb}
\usepackage{upgreek}
\usepackage{float}
\usepackage{subfigure}
\usepackage[dvipsnames]{xcolor}
\usepackage[normalem]{ulem}
\graphicspath{{./figures/},{./figures_eps_r/}}
\newlength{\singlecolumnwidth}
\usepackage{tikz}

\newcommand{\Fl}{\mathrm{Fl}}
\newcommand{\Flh}{\mathrm{Fl}_h}

\newcommand{\zf}{\zeta_f^\parallel}
\newcommand{\zfr}{\zeta_{f,r}^\parallel}

\usepackage[bookmarksnumbered=true,breaklinks=true,colorlinks,citecolor=blue,linkcolor=blue]{hyperref}
\setcitestyle{square}

\makeatletter

\newcommand{\numtoRoman}[1]{\expandafter\@slowromancap\romannumeral #1@}

\makeatother

\usepackage{verbatim}
\usepackage{ulem}
\usepackage{color,soul}

\newcommand{\revision}[1]{\textcolor{black}{#1}}
\newcommand{\redtext}{\color{black}}

\newcommand{\el}{\mathrm{el}}
\newcommand{\mot}{\mathrm{mot}}
\newcommand{\ster}{\mathrm{ster}}
\newcommand{\eq}{\mathrm{eq}}

\newcommand{\C}{\mathrm{C}}

\begin{document}
\title{Impacts of multiflagellarity on stability and speed of bacterial locomotion}
\author{Frank T.~M.~Nguyen}
\affiliation{%
Department of Chemical and Biological Engineering\\
University of Wisconsin-Madison, Madison, WI 53706-1691
}%
\author{Michael D.~Graham}\email[Corresponding author.  E-mail:  ]{mdgraham@wisc.edu}
\affiliation{%
Department of Chemical and Biological Engineering\\
University of Wisconsin-Madison, Madison, WI 53706-1691
}%

\date{\today}

\begin{abstract}
\revision{Trajectories and conformations of uni- and multiflagellar bacteria are studied with a coarse-grained model of a cell comprised of elastic flagella connected to a cell body. The elasticities of both the hook protein (connecting cell body and flagellum) and flagella are varied. Flexibility plays contrasting roles for uni- and multiflagellar swimmers. For a uniflagellar swimmer, hook and/or flagellar buckling occurs above a critical flexibility relative to the torque exerted by the flagellar motor.  Addition of a second flagellum greatly expands the parameter regime of stable locomotion, because flexible hooks that would lead to buckling instability in the uniflagellar case provide the flexibility required for flagellar bundling in the biflagellar case. Similar observations hold for tri- and quadriflagellar swimmers. Indeed the stability regimes for uni- and quadriflagellar swimming are virtually inverted -- to a first approximation what is stable in one case is unstable in the other.   Swimming speed is also examined: it increases very weakly with number of flagella and a simple theory is developed that explains this observation.}

\end{abstract}

\maketitle
\section{Introduction}

A flagellated bacterium swims through viscous fluid by rotating its flagellum or flagella to generate propulsion on its body. The bacterial flagellum is a helical elastic filament connected to an embedded motor in the cell wall via a small elastic hook. A vast array of flagellar morphologies exists amongst different organisms, and studies have directly tied these morphologies to locomotion \cite{Lauga2009, Lauga2016}. For bacterial flagella, the bending stiffness $K_B$ is typically between 1-10 pN $\mu$m$^2$ \cite{Darnton:2007ct,Jawed:2015vw,Son:2013dh}, For bacterial hooks, the bending stiffness $K_{Bh}$ varies widely between species, ranging from 0.2 pN $\mu$m$^2$ for monotrichous (uniflagellar) bacteria to $10^{-4}$ pN $\mu$m$^2$ for peritrichous multiflagellar bacteria \cite{Sen2004, Son:2013dh}. The stark disparity in flexibility between monotrichous and peritrichous bacteria is certainly connected to how they swim.  \revision{The present study examines the contrasting roles of elasticity for uniflagellar and multiflagellar locomotion.  For unflagellar swimmers that are pushed through fluid by their flagellum, stiffness is required for stability. On the other hand, for multiflagellar swimmers whose flagella are anchored at random positions on the cell body, substantial flexibility, especially of the hook, is required for flagella to assemble into the bundles that propel the cell.   Using a coarse-grained micromechanical model, we characterize trajectories, conformations and velocities of swimming uni- and multiflagellar bacteria over a wide range of parameters to elucidate this dichotomy.}

Most bacteria ($>$90\%) are \textit{monotrichous} or uniflagellar, consisting of only one cell body with one attached flagellum \cite{Leifson1964}, e.g. \textit{Vibrio alginolyticus}, \textit{Rhodobacter sphaeroides}, and \textit{Caulobacter crescentus} \cite{Lauga2009}.  Despite their simplicity, uniflagellar locomotion exhibits both normal (straight) trajectories and more complex ones, because these swimmers are able to exploit elastic instabilities in the hook and filament \cite{Lauga2016,Son:2013dh}. \revision{A cell pushed from behind by its flagellum is under compression; the flagellum is pushing forward on the body and the drag from the fluid is pushing backward. In this case the hook may be subject to a buckling instability; this will lead to misalignment between cell body and flagellum, resulting in changes in swimming direction and curved trajectories. 
Specifically, Son et al.~\cite{Son:2013dh} attribute  ``flicks" of the flagellum and the corresponding reorientation in  trajectories of \textit{V. alginolyticus} to buckling of the hook protein. This reorientation capability is crucial to chemotaxis for this species, as it allows a uniflagellar swimmer to effectively undergo run-and-tumble-like trajectories.}  Hook buckling is also 
\revision{responsible for} the helical trajectories observed for \textit{C. crescentus} \cite{Liu:2014ei}.  While a rotating elastic flagellum can also buckle \cite{Jawed:2015vw} under the compression associated with the thrust it produces, this feature does not seem to be observed in normal bacterial locomotion \cite{Lauga2016}. Rather, flagellar elasticity is associated with polymorphic phase transformations in a flagellum. For example, in \textit{Shewanella putrefaciens}, this transformation helps its polar flagellum wrap around the cell body to escape confinement via a corkscrew motion \cite{Kuhn:2017ix}.

For multiflagellar bacteria, the \textit{peritrichous} morphology, where many flagella are attached at random points on the body surface, is quite common. Peritrichous swimmers have different swimming mechanisms than the monotrichous ones. For instance, in the classic run and tumble example of \textit{E. coli}, the elastic flagella bundle to swim straight and unbundle to change direction \cite{Berg2003,Darnton:2007ct}. Flagellar bundling is a common phenomenon across peritrichous organisms, and is exhibited in other species such as \textit{Bacillus subtilis} and \textit{Salmonella typhimurium} \cite{Hyon:2012hj}. For bundling to occur, the hooks must be flexible to allow the flagella to assemble into a bundle -- with stiff hooks, the flagella would just stick out from the body as they rotate, in which case bundling cannot occur\cite{Darnton:2007ct}.  Interestingly, multiflagellar swimmers do not necessarily swim much faster than their uniflagellar counterparts \cite{Darnton:2007ct} -- \revision{therefore, multiflagellar swimming probably confers other advantages, such as 
the ability to navigate complex geometries, or locomotion on surfaces or in complex media as noted in Refs. \cite{Rusconi:2014eh,Quelas:2016iq}.}


\revision{
A number of aspects of the role of elasticity in the stability and dynamics of bacterial locomotion with flagella have been studied theoretically and computationally. 
With regard to uniflagellar swimming, Shum and Gaffney \cite{Shum:2012hy} studied the stability of uniflagellar locomotion using a Kirchoff rod model of a flexible hook along with a boundary integral treatment of a rigid cell body and flagellum.  They found that an overly flexible hook displayed a buckling instability that precluded stable swimming. 
Nguyen and Graham \cite{Nguyen:2017jt} made similar observations using much simpler models that neglected hydrodynamic interactions and treated the hook as a simple spring coupling.
Jabbarzaheh and Fu \cite{Jabbarzadeh2018} used a model that allowed for both hook and flagellar flexibility to shed light on the ``flicking" motion experimentally observed by Son et al. \cite{Son:2013dh}.
Vogel and Stark \cite{Vogel:2012bm} mapped out the a bifurcation diagram for the dynamics of an anchored rotating elastic flagellum. (The mathematical model we develop below is closely related to theirs.) When the flagellum is rotating so as push on the anchor point (the situation for a flagellum pushing a cell body forward), there is a regime at low torque for which the flagellar axis remains parallel to the direction of the imposed torque. At higher torque there is a buckling instability leading to bending of the flagellum and a stable oscillation in the thrust. At higher torques, this limit cycle becomes unstable. Jawed et al.~\cite{Jawed:2015vw} observe similar results in their simulations and experiments.  If the torque is reversed, corresponding to pulling, Vogel and Stark find that the straight helix is unstable. This result may be related to the experimental and computational observations of K\"uhn et al. \cite{Kuhn:2017ix}; they find experimentally and corroborate with simulations that \emph{Shewanella putrefaciens} displays a phenomenon during a pulling phase of locomotion in which the flagellum, instead of extending from the body, wraps around it, generating a very unusual mode of locomotion. 
}

\revision{Multiflagellar locomotion with bundles has also received some attention. Formation of a robust bundle requires the flagella to approach one another and then intertwine as they rotate. To some extent, the bundling process can be induced simply through the counterrotation of the cell body to which the flagellar are attached \cite{Powers:2002p4790}.  Even if the cell is prevented from rotating (or equivalently, the flagellar motors are anchored in place), it has been found experimentally \cite{Kim2004} and computationally \cite{Janssen:2011ex, Reigh:2012fq} that the fluid motion generated by flagella as they rotated (i.e. the hydrodynamic interactions (HI) between them) are sufficient to cause flexible to begin to attract and intertwine with one another. For the flagella to intertwine persistently, they must synchronize; Qian et al.~\cite{Qian:2009p4716}  and Janssen and Graham \cite{Janssen:2011ex} have shown, with models of rotating ``paddles'' or flagella, that synchronization, and specifically phase locking, arises naturally under driving at constant motor torque.  Reigh et al.~\cite{Reigh:2012fq,Reigh:2013ga} show that synchronization is robust under variations in torque between different flagellar motors. Watari and Larson \cite{Watari:2010p6425} developed a very coarse-grained bead-spring assembly model of a multiflagellar bacterium that incorporates flagellar polymorphism under counter-rotation; the model is able to reproduce the classical run-and-tumble motion exhibited by \emph{E.~coli}. More detailed analysis of polymorhism has been performed by Vogel and Stark \cite{Vogel:2013ht}. Adhyapak and Stark \cite{Adhyapak:2015jk} performed a detailed computational study of the dynamics of flagellar bundle formation, contrasting the cases of anchored flagella and flagella on a free-swimming cell. Focusing on a biflagellar cell, they describe processes of ``zipping'' and ''entanglement'' and how they contribute the overall bundling process.  
Finally, in the work most closely related to the present study, Riley et al.~\cite{Riley2018} describe the dynamics of a cell with multiple rigid flagella, modeled as tapered helices, connected by flexible hooks, modeled as torsion springs, to a rigid prolate spheroidal body. Hydrodynamic interactions between body and flagella are neglected. The flagella are driven at constant angular velocity and are symmetrically arranged on the body in opposition to one another so that a symmetry-breaking instability must arise for swimming to occur. Indeed, if the flagellar are rotating so as to push on the body, the hook is under compression and if it is sufficiently flexible, an instability closely related to that found in the uniflagellar case by Shum and Gaffney \cite{Shum:2012hy} and Nguyen and Graham \cite{Nguyen:2017jt} is observed. Once symmetry is broken, the flagella are no longer exerting forces in opposing directions on the body body and the cell will swim.    
}

\revision{The present work comprises a comprehensive and systematic study of uni- and multiflagellar swimming over a wide range of elastic properties of the hook and flagellum. A coarse-grained model of a spherical cell with one or more elastic flagella is described, and we map out the dependence of the swimming mode on elastic parameters for cells with flagella anchored on the cell surface at one to four corners of a regular tetrahedron inscribed in the sphere.  For a uniflagellar swimmer, hook and/or flagellar buckling occurs above a critical flexibility relative to the torque exerted by the flagellar motor.  Addition of a second flagellum greatly expands the parameter regime of stable locomotion, because flexible hooks that would lead to buckling instability in the uniflagellar case provide the flexibility required for flagellar bundling in the biflagellar case. Similar observations hold for tri- and quadriflagellar swimmers. Indeed the stability regimes for uni- and quadriflagellar swimming are virtually inverted -- to a first approximation what is stable in one case is unstable in the other.  We also show that aside from symmetry, the tetrahedral arrangement is not special -- for sufficiently flexible flagella and hooks, quadriflagellar bundling and swimming is fairly robust with respect to flagellar arrangement. Swimming speed is also examined: it increases very weakly with number of flagella and a simple theory is developed that explains this observation.}


\section{Model and discretization}

\begin{figure}
	\begin{minipage}{0.48\textwidth}
	\includegraphics[width=\linewidth]{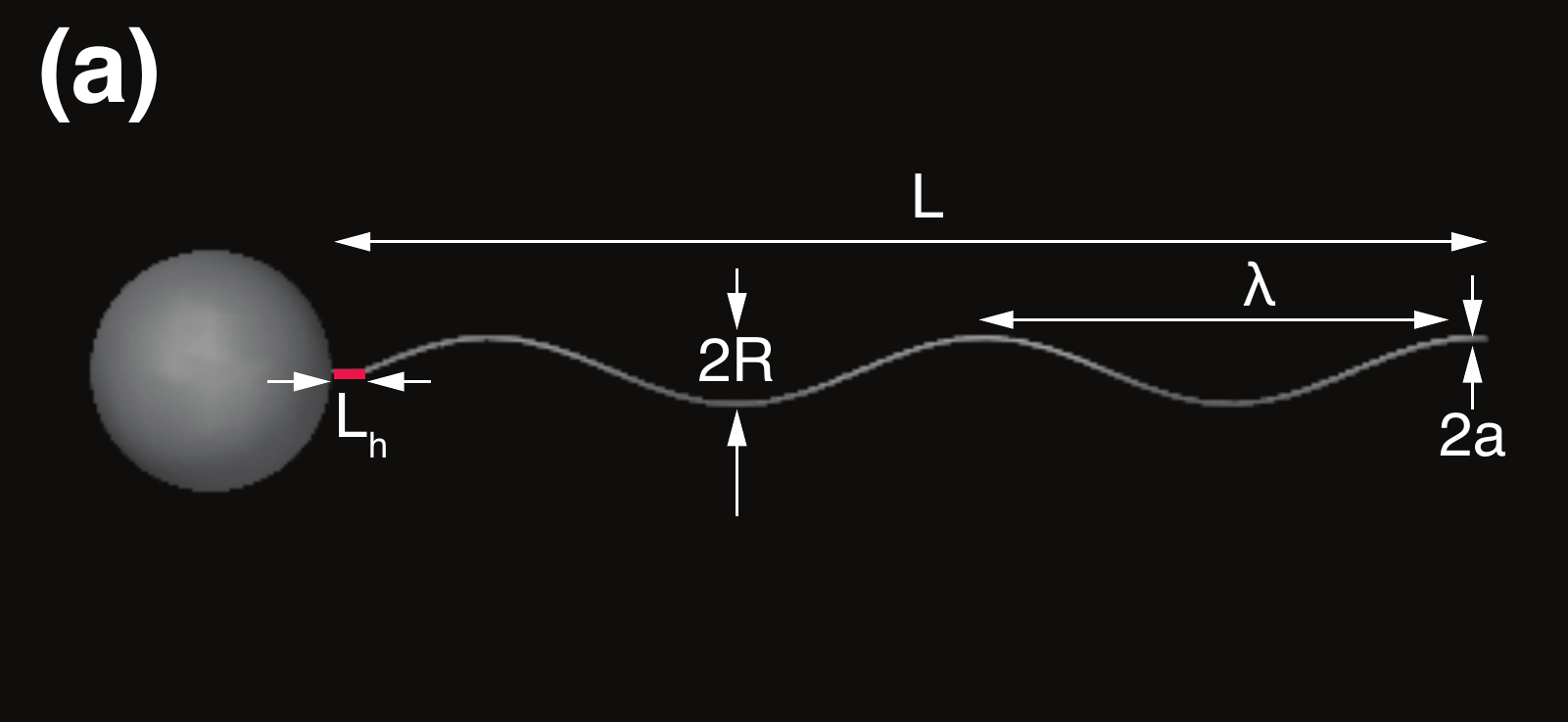}
	\end{minipage}
	\begin{minipage}{0.48\textwidth}
		\includegraphics[height=1.575in]{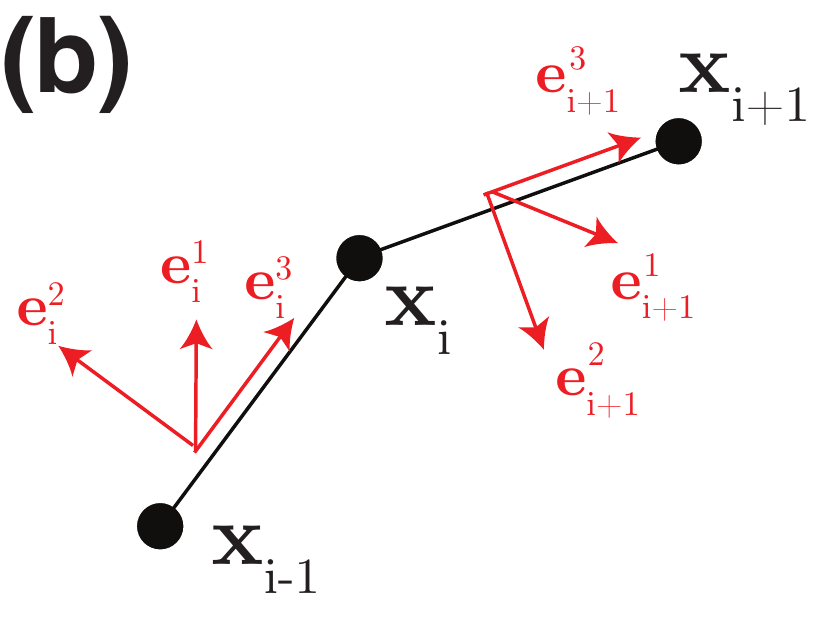} \\
		\includegraphics[height=1.675in]{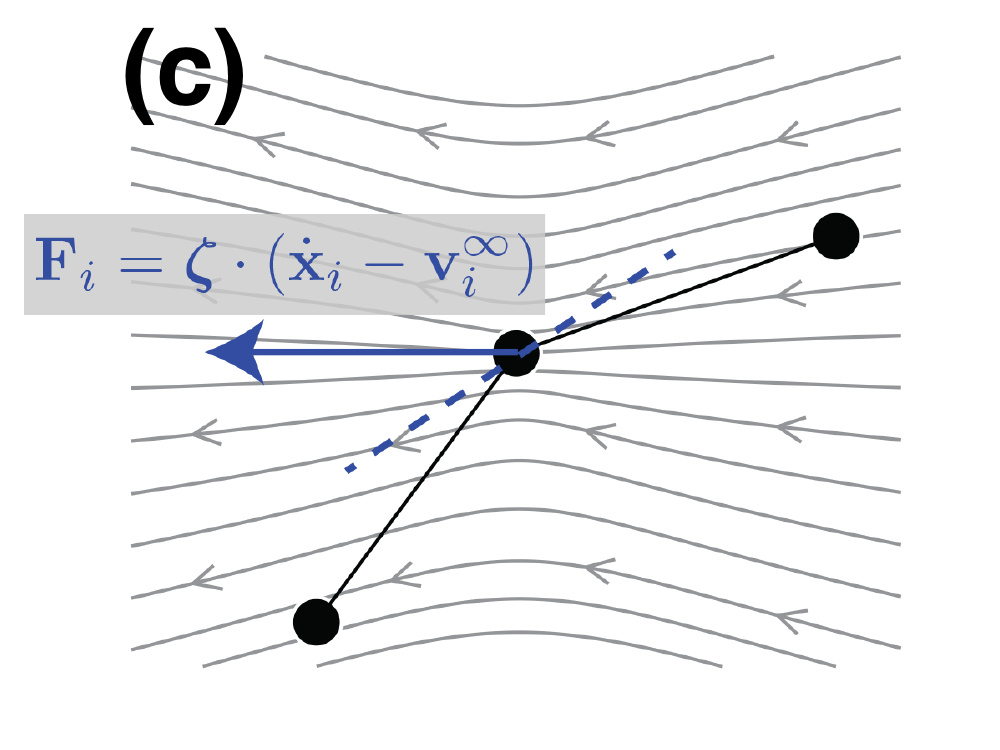}
	\end{minipage}
	\caption{(a) Model swimmer with flagellum. (b) Discretization of flagellum into nodes $\mathbf{x}_i$ and edges $\mathbf{e}_i$. (c) Flow field generated by a regularized point force (gray lines). The dashed blue line shows the approximation of the local tangent vector at node $\mathbf{x}_i$.} 
	\label{fig_model3d}
\end{figure}

\subsection{Physical description}
Our model swimmer, depicted in Fig. \ref{fig_model3d}, consists of a rigid spherical cell body of radius $R_b$ connected to $N$ right-handed flagella \revision{of length $L$}, each via a flexible hook of length $L_h$. 
Each flagellum is an inextensible, elastic helix; \revision{in multiflagellar cases all flagella will have the same mechanical and geometric properties.}
  In addition to length $L$, the equilibrium helix geometry is described by radius $R$, pitch $\lambda$, and filament radius $a$. The elasticity of the flagella and hooks are described by flexural rigidities $K_B$ and $K_{Bh}$, respectively. \revision{Further details on hook mechanics and geometry are described in Section \ref{sec:elasticity}.} During swimming,  motors embedded within the body surface exert torques of constant magnitude $T$ on the hooks, which in turn transmit these torques to their respective flagella. \revision{The choice of constant torque rather than constant angular velocity is important for two reasons. First, experiments indicate that, to a good approximation, flagellar motors operate at constant torque over a wide range of loads \cite{CHEN:2000bo,Yuan:2010dp}. Second, constant-torque driving is robust (which is perhaps why nature uses it). Specifically, synchronization of flagella driven by constant-torque motors is robust with respect to initial phase difference of the motors \cite{Janssen:2011ex} and against mismatch of the motor torques \cite{Reichert:2005p48} By contrast, for multiple motors running at constant angular velocity, the phase relation between the motors is set for all time by the initial condition and only particular sets of motor phases will allow multiple flagella to mesh into a bundle. Furthermore, for motors that run at constant but different speeds, the phase difference increases with time, precluding any kind of quasisteady bundling process.}

\revision{A critical issue in this work will be buckling phenomena associated with the hooks and flagella. For a simple Euler beam of length $L$ with fixed ends, the critical buckling force is $\pi^2 K_B/L^2$ under compression and the critical buckling torque is $2\pi K_B/L$ under torsion    \cite{timoshenko1961theory}}. Motivated by this observation, we characterize the flexibility of the flagella and hook by dimensionless \emph{flexibility numbers} $\Fl$ and $\Fl_h$, defined as
\begin{equation}
	\Fl = \frac{T}{K_B/L}, \; \; \Flh = \frac{T}{K_{Bh}/L_h}.
	\label{eq:flex}
\end{equation}
We note that Vogel and Stark~\cite{Vogel:2012bm} describe an analysis of the bending elasticity of a helical filament that provides a geometry-dependent correction to the simple estimate $K_B$.

For a bacterial flagellum, $K_B$ is typically in the range 1-10 pN $\mu$m$^2$ \cite{Darnton:2007ct,Jawed:2015vw,Son:2013dh}, $T\sim 0.5-3$ pN $\mu$m \cite{Darnton:2007ct}, and $L\sim 3-10$ $\mu$m \cite{Rodenborn:2013iy}. This translates to $\Fl$ between 0.3 and 30. \revision{We are not aware of organisms with higher values of $\Fl$. For the peritrichous species \emph{E.~coli} and \textit{S. typhimurium}, $\Fl\approx 2.5$, and for the monotrichous \textit{V. alginolyticus}, $\Fl\sim 0.2-1.6$ \cite{Darnton:2007ct, Son:2013dh}. In our results below we consider $\Fl$ from $1.6$ to $6.5$. Bacterial hooks have length $L_h\approx 100$ nm for uniflagellar organisms and 50-80 nm for peritrichous organisms, with very wide ranges of $K_{Bh}$, ranging from 0.2 pN $\mu$m$^2$ (upper uniflagellar estimate) to $10^{-4}$ pN $\mu$m$^2$ (peritrichous) \cite{Sen2004, Son:2013dh}. From this data, an approximate range for $\Flh$ is as low as 0.2 for uniflagellar organisms and as high as 1000 for peritrichous. 
Our simulations explore the range $10^{-2}\leq\Flh\leq 10^2$.}

\subsection{\revision{Geometry and discretization}}

The cell body is a rigid sphere with center of mass position $\mathbf{x}_b$ and orientation $\mathbf{q}_b$, a unit quaternion (whose components are Euler parameters). Initially, we align the body orientation with the laboratory frame so that $\mathbf{q}_b = [1 \; 0 \;0 \; 0]^\mathrm{T}$. We select anchor points on the body surface where at each point we attach a hook and flagellum.
 
Each flagellum is discretized into $M-1$ connected straight rods of radius $a$ which we call \textit{edges}. The edge connected to the body at the anchor point is the hook, which has length $L_h$. At rest it extends normally from the body. 
The remaining edges are flagellar segments each of length $l$, constituting a plain helix directly attached to the hook. \revision{With this construction, edge 1 (the hook) and edge 2 (the first flagellar segment) are joined with an angle given by the pitch angle $\tan^{-1}\frac{\lambda}{2\pi R}$ of the filament, while the other angles are determined by the helix geometry and degree of discretization to form the discretized helix.}  The points connecting each edge, including endpoints, are called \textit{nodes}. Thus, flagellum $j$ has $M$ nodes  $\mathbf{x}_i^j$ with $i\in [0,\,  M-1], j\in [1,N]$.  Node $0$ denotes the anchor point attached to the body surface. A schematic is shown in Fig. \ref{fig_model3d}(b).

To simplify notation, we will generally drop the superscript representing which flagellum a node is on.  For the edge $i$ between nodes $i-1$ and $i$ (of flagellum $j$), we assign a local orthonormal coordinate system or \textit{triad} $\{ \mathbf{e}^1_i, \mathbf{e}^2_i, \mathbf{e}^3_i \}  $, where $\mathbf{e}_i^3$ is exactly the edge defined by:
 \begin{subequations}
 	\begin{equation}
 	\mathbf{e}_i^3 = \frac{\mathbf{x}_i - \mathbf{x}_{i-1}}{|\mathbf{x}_i - \mathbf{x}_{i-1}|} , \; \; i \in [1, \, M-1],
 	\end{equation}
 	\begin{equation}
 	\mathbf{e}_0^3 = \frac{\mathbf{x}_0 - \mathbf{x}_{b}}{R_b}.
 	\end{equation}
 	\label{eq:e3}
 \end{subequations}
Note that \revision{$\mathbf{e}_0^3$ is simply a unit vector pointing normally outward from the cell body at the the position of the anchor point. It defines the axis of the motor driving the corresponding flagellum. Similarly,} $\mathbf{e}_1^3$ defines the hook orientation, which at rest is parallel to the flagellar axis. The remaining triad components are defined as follows:
\begin{subequations}
	\begin{equation}
		\mathbf{e}_i^1 = \frac{\mathbf{e}_{i-1}^3 \times \mathbf{e}_i^3}{|\mathbf{e}_{i-1}^3 \times \mathbf{e}_i^3|},
	\end{equation}
	\begin{equation}
		\mathbf{e}_i^2 = \mathbf{e}_i^3 \times \mathbf{e}_i^1.
	\end{equation}
	\label{eq:e1e2}
\end{subequations}
\revision{Eqs. \ref{eq:e3} and \ref{eq:e1e2} are used only to initialize the flagellar helix, after which point we integrate the triads directly rather than reconstructing them (see Appendix \ref{app:numerics}). We do not use Eqs. \ref{eq:e3} and \ref{eq:e1e2} to initialize the edges $\mathbf{e}_0^3$ and $\mathbf{e}_1^3$.} \revision{We will not need to keep track of $\mathbf{e}_0^1$ or $\mathbf{e}_0^2$ because they play no role in the dynamics.} 

Given the triads $\{ \mathbf{e}^1_i, \mathbf{e}^2_i, \mathbf{e}^3_i \}  $, we define a bend angle $\theta_i$ and twist angle $\varphi_i$ for node $i$ that transform triad $i$ to triad $i+1$ \cite{Reichert2006}. For $i=0,1,2,\ldots,M-2$:
\begin{subequations}
	\begin{equation}
		\theta_i = \arccos \left( \mathbf{e}_i^3 \cdot \mathbf{e}_{i+1}^3 \right),
	\end{equation}
	\begin{equation}
		\varphi_i = \arctan \frac{\tilde{\mathbf{e}}_i^1 \cdot \mathbf{e}_i^2}{\mathbf{e}_i^1 \cdot \tilde{\mathbf{e}}_i^1},
	\end{equation}
	\begin{equation}
		\tilde{\mathbf{e}}_i^1 = \left[ \mathbf{nn}  + \cos \theta_i  \left( \boldsymbol{\updelta} - \mathbf{nn} \right) \right] \cdot \mathbf{e}_{i+1}^1 - \sin\theta_i \left( \mathbf{n} \times \mathbf{e}_i^1 \right),
	\end{equation}
	\begin{equation}
		\mathbf{n} = \mathbf{e}_i^3 \times \mathbf{e}_{i+1}^3 / \sin\theta_i = \mathbf{e}_{i+1}^1.
	\end{equation}
\end{subequations}
Here $\boldsymbol{\updelta}$ is the identity tensor, and $\tilde{\mathbf{e}}_i^1$ is an intermediate twist-free rotation of $\mathbf{e}_i^1$. \revision{(Actually the twist angle $\phi_0$, i.e.~the twist of the hook relative to the motor, plays no role in the dynamics and need never be computed.)}
With these quantities we define the generalized curvature, $\boldsymbol{\Omega}_i$:
\begin{subequations}
	\begin{equation}
		\Omega_i^1 = - \frac{\theta_i}{\sin\theta_i} \mathbf{e}_i^2 \cdot \mathbf{e}_{i+1}^3,
	\end{equation}
	\begin{equation}
	\Omega_i^2 = + \frac{\theta_i}{\sin\theta_i} \mathbf{e}_i^1 \cdot \mathbf{e}_{i+1}^3,
	\end{equation}
	\begin{equation}
	\Omega_i^3 = \varphi_i.
	\end{equation}
	\label{eq:curv}
\end{subequations}
Eq. \ref{eq:curv} is also used to determine the equilibrium curvature, $\boldsymbol{\Omega}_{i,\eq}$.  With $\boldsymbol{\Omega}_i$, we can describe the flagellar conformation needed to characterize elasticity and its evolution during motion. In Section \ref{sec:elasticity} we show how the elastic energy of the system is written in terms of these quantities.

\revision{To conclude this section we summarize the degrees of freedom required to describe the position and conformation of a cell. The quantities $\mathbf{x}_b$ and $\mathbf{q}_b$ determine the position and orientation of the cell body. Each flagellum is described by the positions $\mathbf{x}_i$ of each node $i=0,1,\ldots,M-1$, and the orientation vectors $\mathbf{e}_i^1$ and $\mathbf{e}_i^2$ of each edge. Incorporating the unit length constraint on $\mathbf{q}_b$,
\begin{equation}
\mathbf{q}_b \cdot \mathbf{q}_b - 1 = 0,
\label{eq:const1}
\end{equation}
 yields $6+9MN$ degrees of freedom.}

\section{Equations of motion}

\revision{
\subsection{Overview}
The cell moves through incompressible Newtonian fluid. Because of the small scale of a bacterial cell, we neglect the inertia of both the fluid and cell, so the sums of forces and torques on the cell body and on each element of each flagellum sum to zero.  There are no externally-imposed forces or torques so the swimmer is also force- and torque-free overall.  The discrete formalism we use is essentially the same used in Refs.~\cite{Bergou2008, Reichert2006, Vogel:2010gc}; we impose force balances on each node to determine their translational motions, and torque balances on each edge to determine their rotations. The force and torque balances on the body are 
\begin{equation}
	\mathbf{F}_b^{\mathrm{D}} + \mathbf{F}_b^\el + \mathbf{F}_b^\ster + 
	\mathbf{F}_b^\C  =  \mathbf{0},
	\label{eq:force_balance_body}
\end{equation}
\begin{equation}
	\mathbf{T}_b^{\mathrm{D}} + \mathbf{T}_b^\el + \mathbf{T}_b^\mot + \mathbf{T}_b^\C  = \mathbf{0},
	\label{eq:torque_balance_body}
\end{equation}
where superscripts indicate hydrodynamic drag (D), elasticity (el), steric interactions (ster), application of motor torque (mot), and constraints (C).
Similarly, there is a force balance on each node $i$ of each flagellum $j$:
\begin{equation}
	\mathbf{F}_i^{\mathrm{D},j} + \mathbf{F}_i^{\el,j} + \mathbf{F}_i^{\ster,j} + \mathbf{F}_i^{\mot,j} + \mathbf{F}_i^{\C,j}  =  \mathbf{0}.
	\label{eq:force_balance_node}
\end{equation}
Constraint forces and the corresponding moments arise from the connectivity requirement relating the cell orientation to the anchor positions,
\begin{equation}
\mathbf{q}_b \mathbf{x}_0^{(bj)} \mathbf{q}_b^{-1} - (\mathbf{x}_0^j - \mathbf{x}_b) = \mathbf{0},
\label{eq:const2}
\end{equation}
and the constant length constraints for the hooks and flagellar segments,
\begin{equation}
(\mathbf{x}_{i+1}^j - \mathbf{x}_i^j) \cdot (\mathbf{x}_{i+1}^j - \mathbf{x}_i^j)  - l_i^2 = 0.
\label{eq:const3}
\end{equation}
\revision{We set $l_i$ to be $L_h$ for a hook and $l$ otherwise}. In Eq. \ref{eq:const2}, $\mathbf{x}_0^{(bj)}$ is a constant vector denoting the anchor point of flagellum $j$ in the body-fixed frame of reference, and the conjugation operation $\mathbf{q}_b \mathbf{x}_0^{(bj)} \mathbf{q}_b^{-1}$ rotates $\mathbf{x}_0^{(bj)}$  to the laboratory frame. With this formulation, the propulsive force exerted by the flagella on the cell body as they rotate is incorporated in the constraint forces that enforce the connectivity condition, Eq.~\ref{eq:const2} (cf.~\cite{Nguyen:2017jt}).  
All steric forces act between either different flagellar nodes or between a nodes and the center of the body, so there is no steric contribution to the torque balances on the body or nodes. Additionally, all bending torques on edges are resolved into forces on nodes, so the only component of the torque balance that needs to be treated explicitly is for twist:
\begin{equation}
	\left(\mathbf{T}_i^{\mathrm{D}} + \mathbf{T}_i^\el + \mathbf{T}_i^\mot\right)\cdot\mathbf{e}_i^3=0. \label{eq:torque_balance_edge}
\end{equation}
This equation, along with kinematic equations relating node velocities to edge angular velocities, completes the set of governing equations -- these are given explicitly as Eqs.~\ref{eq:updatees}, \ref{eq:k1} and \ref{eq:k2} in Appendix \ref{app:numerics}. In summary, $6$ equations arise from the force and torque balances for the body, $3MN$ from the force balances on each node, and another $6MN$ from the torque balances and kinematic equations for the edges for a total of $6+9MN$ equations that are solved at each time step for the unknowns.
}

\revision{In the following subsections we detail the evaluation of the various forces and torques and summarize the numerical algorithm for time-evolution of the governing equations; details of the latter are given in Appendix \ref{app:numerics}.}

\subsection{Hydrodynamics}


\revision{The hydrodynamic treatment of the flagella used here is a discretized approximate version of slender body theory \cite{Tornberg2004}; each discretized segment (``rod'') of a flagellum experiences an anisotropic Stokes drag force; the (equal and opposite) force exerted by the segment as it translates generates a fluid motion that is felt by all the other flagellar segments and the body.}
Each node on a flagellum translates with velocity $\dot{\mathbf{x}}_i = \mathbf{v}_i$. As it moves, the surrounding fluid exerts a Stokes law drag force on it. This drag force $\mathbf{F}_i^{\mathrm{D}}$ has three contributions: a local contribution from the motion of node $\mathbf{x}_i$ itself, and far-field contributions induced by the motion of all other flagellar nodes ($\mathbf{v}_{i,\infty}^f$) and the body ($\mathbf{v}_{i,\infty}^b$). In this case
\begin{equation}
\mathbf{F}_i^\mathrm{D} = \boldsymbol{\zeta}_i \cdot \left[\left( \mathbf{v}_{i,\infty}^f + \mathbf{v}_{i, \infty}^b - \mathbf{v}_i  \right) \right],
\label{eq:force_drag_flag}
\end{equation}
where $\boldsymbol{\zeta}_i$ is an anisotropic friction coefficient for a slender rigid rod of aspect ratio $l/a$:
\begin{equation}
\boldsymbol{\zeta}_i = \zeta_\perp \boldsymbol{\delta} + (\zeta_\parallel - \zeta_\perp) \mathbf{t}_i \mathbf{t}_i,
\label{eq:drag_tensor}
\end{equation}
and $\zeta_\perp$ and $\zeta_\parallel$ are the scalar normal and tangential friction coefficients for rods of the given geometry \cite{Rodenborn:2013iy}. We take the axis of the rod at node $i$ to be oriented with the average orientation of edges $i$ and $i+1$: $\mathbf{t}_i = \frac{1}{2} (\mathbf{e}_i^3 + \mathbf{e}_{i+1}^3)$. Using the Stokes drag for a rod rather than a sphere at each node automatically incorporates the anisotropy of drag that leads to flagellar locomotion and allows for a coarser discretization than would be necessary if we had used spheres \cite{Janssen:2011ex}. \revision{For the helical geometry used here, we have found that the results are insensitive to $l$ as long as at least 10 rods per helical wavelength are used.}

\revision{Traditional slender body theory does not account for the fluid frictional torque associated with rotation of flagellar segments around their local axes. Because the rotational friction coefficient $\zeta^r=4\pi\eta a^2 l$ scales as $a^2$, this drag torque is very small. Nevertheless, for a flexible filament in the absence of these forces, the local twisting angles are in quasi-steady state given the distribution forces along the filaments, and the forces and torques exerted at its ends. Such a quasistatic description is used, for example, by Bergou et al.~\cite{Bergou2008}, resulting in a system of differential-algebraic equations. Here, however, we have elected to retain the rotational friction in the torque balance for each segment to obtain dynamical equations for the rotation of each segment rather than algebraic ones. Adhyapak and Stark \cite{Adhyapak:2015jk} also take this approach. In this case, each rod experiences a drag torque
\begin{equation}
\mathbf{T}_i^\mathrm{D} = - \zeta_i^r \omega_i \mathbf{e}_i^3.
\end{equation}
}
%

The velocity field $\mathbf{v}_{i,\infty}^f$ in Eq.~\ref{eq:force_drag_flag} results in hydrodynamic interactions between the flagella. To compute this field, we treat each node as a regularized point force acting on the fluid, as illustrated in Fig. \ref{fig_model3d}(c). 
 The flagellar flow field $\mathbf{v}_{i,\infty}^f$ experienced by node $i$ is obtained by summing the flows induced by all other flagellar nodes:
\begin{equation}
\mathbf{v}_{i,\infty}^f =  \sum_{i \neq j}  \mathbf{S}_\xi(\mathbf{x}_i, \mathbf{x}_j) \cdot \mathbf{F}_j,
\label{eq:flagellar_flow}
\end{equation}
The tensor $\mathbf{S}_\xi$ is the regularized Stokeslet defined in Appendix \ref{App:AppendixA}, and $\mathbf{F}_j = -\mathbf{F}_j^\mathrm{D}$ is the force exerted \textit{on} the fluid \textit{by} flagellar node $j$. 
%

To obtain the body induced flow $\mathbf{v}_{i,\infty}^b$, we use the Stokes flow solution for a sphere experiencing force $\mathbf{F}_b$ and torque $\mathbf{T}_b$ moving through stationary fluid:
\begin{equation}
\mathbf{v}_{i,\infty}^b = \left( 1 + \frac{R_b^2}{6} \nabla^2 \right) \mathbf{S}(\mathbf{x}_i - \mathbf{x}_b)\cdot\mathbf{F}_b  \, + \, \frac{R_b^3}{\zeta_b^r} \mathbf{R} \cdot  \mathbf{T}_b,
\label{eq:flow_sphere}
\end{equation}
where $\mathbf{S}$ is the standard Stokeslet tensor and $\mathbf{R}$ the rotlet tensor, both defined in Appendix \ref{App:AppendixA}. The scalar $\zeta_b^r = 8 \pi \eta R_b^3$ is the rotational friction coefficient for the sphere. With regard to Eq. \ref{eq:flow_sphere}, $\mathbf{F}_b = -\mathbf{F}_b^\mathrm{D}$ and $\mathbf{T}_b = -\mathbf{T}_b^\mathrm{D}$, and thus we require the body drag to fully calculate flagellar hydrodynamics.

The body moves with velocity $\mathbf{v}_b$ and angular velocity $\boldsymbol{\omega}_b$. We write the hydrodynamic force and torque on the sphere using Fax\'en's laws to account for body-flagella HI as was done by Tam and Hosoi \cite{Tam:2011iw}:
\begin{equation}
\mathbf{F}_b^\mathrm{D} = \zeta_b \left[ -\mathbf{v}_b +  \left. \left( 1 + \frac{R_b^2}{6} \nabla^2  \right) \mathbf{v}_{\infty}^f  \right|_{\mathbf{x}_b} \right],
\label{eq:force_drag_body}
\end{equation}
\begin{equation}
\mathbf{T}_b^\mathrm{D} = \zeta_b^r \left[- \boldsymbol{\omega}_b + \left. \frac{1}{2} \left( \nabla \times \mathbf{v}_\infty^f  \right) \right|_{\mathbf{x}_b} \right].
\label{eq:torque_drag_body}
\end{equation}
Here $\zeta_b = 6 \pi \eta R_b$ is the translational drag on a sphere. We relate $\mathbf{q}_b$ to $\boldsymbol{\omega}_b$ via quaternion multiplication:
\begin{equation}
\arraycolsep=1.4pt\def\arraystretch{0.65}
\dot{\mathbf{q}_b} = \frac{1}{2}\left[ \begin{array}{c} 0 \\ \boldsymbol{\omega}_b \end{array} \right] \mathbf{q}_b
\label{eq:quaternion}
\end{equation}
The flagellar flow $\mathbf{v}_\infty^f$ is written generally for a point $\mathbf{x}$ anywhere in the fluid:
\begin{equation}
\mathbf{v}^f_\infty (\mathbf{x}) = \sum_j \mathbf{S}_\xi(\mathbf{x}, \mathbf{x}_j) \cdot \mathbf{F}_j.
\label{eq:flagellar_flow_full}
\end{equation}
Derivatives of $\mathbf{v}_\infty^f$ in Eqs. \ref{eq:force_drag_body} \& \ref{eq:torque_drag_body} are evaluated at the body center $\mathbf{x}_b$.  In contrast to Eq. \ref{eq:flagellar_flow}, all flagellar nodes are included in the summation in Eq. \ref{eq:flagellar_flow_full}.

We conclude this section with a brief discussion of the limitations of this hydrodynamic formulation. Our aim has been to incorporate the primary hydrodynamic interaction effects in a  computationally efficient framework that would allow a broad exploration of parameter space.  The neglect of the potential dipole term in slender body theory introduces errors that scale as $(a/r)^3$; for a bacterial flagellum $a\sim 10-15$ nm, so even on distance scales $r\sim 30$ nm, this error is small. A perhaps more significant error is in treating the flow generated by flagellar segments near the body as free-space Stokeslets: we neglect the presence of a nearby no-slip wall, namely the cell body. As such the present formulation overestimates the flows generated by flagellar segments near the wall. This error is mitigated to some extent by the fact that for long flagella, most of the segments are far from the wall and these will in general dominate the flow field. 
    
\subsection{Elasticity}\label{sec:elasticity}

Having previously defined the discrete generalized curvatures $\boldsymbol{\Omega}_i$ and $\boldsymbol{\Omega}_{i,\eq}$, we use a discrete version of Kirchhoff's classical theory to write the elastic energy, $\mathcal{E}^\el$, of each flagellum as the sum of hook and flagellum contributions:
\revision{
\begin{equation}
\mathcal{E}^\el = \frac{1}{2}\frac{K_{Bh}}{L_h} \left(\theta_{0} - \theta_{0,\mathrm{eq}} \right)^2 + \frac{1}{2} \frac{K_B}{l}   \sum_{i=1}^{M-2} \left[ \sum_{\beta=1}^2 \left(\Omega_i^\beta - \Omega_{i,\mathrm{eq}}^\beta \right)^2 + \Gamma \left( \Omega_i^3 - \Omega_{i,eq}^3 \right)^2 \right],
\label{eq:elastic_energy}
\end{equation}
We set $\theta_{0,\mathrm{eq}} = 0$ (straight hook at equilibrium) and assign no twist penalty to allow for free rotation and counter-rotation between body and flagella. We assume the flagellar filament has a circular cross section so that bending is isotropic. The parameter $\Gamma = K_T/K_B$ is a material property called the twist-to-bend ratio, where $K_T$ is the torsional ridity. For an elastic rod of incompressible material with circular cross section, $\Gamma = 2/3$. However, force-extension experiments with bacterial flagella, Darnton et al. \cite{Darnton:2007p6433} found that $\Gamma = 1$ fit their data quite well, so here we set $\Gamma=1$ as well. Shum and Gaffney \cite{Shum:2012hy} and Adhyapak and Stark \cite{Adhyapak2015} also made this choice.   Spot checks with $\Gamma=2/3$ indicate that the results presented here are robust with regard to the specific value chosen.}

\revision{Our model for the hook is very simple -- in particular it take the twisting rigidity of the hook to be the same as the flagellum. Shum and Gaffney \cite{Shum:2012hy} and Jabbarzaheh and Fu \cite{Jabbarzadeh2018} have used more detailed models treating the hook as a very flexible Kirchoff rod to try to address the relation between hook dynamics and swimming.  A potentially important issue in some cases is what happens when the hook becomes highly twisted and thus displays very nonlinear behavior. We have made no attempt to capture these details.  However, we do find that increasing or decreasing the twist resistance of the hook and flagellum by factor of 2 leads to no substantial differences in the dynamics.}

On each flagellum, derivatives of the elastic energy determine the elastic force on node $i$ and elastic torque on edge $i$: 
\begin{equation}
\mathbf{F}_{i}^\el = -\frac{\partial \mathcal{E}^\el}{\partial \mathbf{x}_{i}},
\label{eq:force_elastic}
\end{equation}
\begin{equation}
\mathbf{T}_{i}^\el =- \frac{\partial \mathcal{E}^\el}{\partial \varphi_{i}} \mathbf{e}_{i}^3 = -T_{i}^\el \mathbf{e}_{i}^3.
\label{eq:torque_elastic}
\end{equation}
The separate treatment of forces and torques in this manner eases the calculation and implementation of twisting and bending deformations on the filament, as noted by Reichert \cite{Reichert2006}. We emphasize that the elastic torque is distributed entirely along the edge direction, $\mathbf{e}_{i}^3$. 

\revision{Finally, the elastic forces and torques on the body result from} elastic forces acting at flagellar anchor points:
\revision{
	\begin{equation}
		\mathbf{F}_b^\el = \sum_\mathrm{flagella} \mathbf{F}_0^\el,
	\end{equation}
}
\begin{equation}
	\mathbf{T}_b^\el = R_b \sum_\mathrm{flagella} \mathbf{e}_0^3 \ \times \mathbf{F}_0^\el,
	\label{eq:torque_elastic_body}
\end{equation}
where the index 0 denotes an anchor point  for a given flagellum.

\subsection{Steric repulsion}

Following Ref. \cite{Adhyapak:2015jk}, we introduce repulsive steric forces when flagellar nodes approach the body and/or other flagellar nodes too closely. A truncated Lennard-Jones potential is used:
\begin{equation}
	U_{LJ} = \frac{F_s \, \sigma}{6} \left[ \left(\frac{\sigma}{r_s}\right)^{12} -  \left( \frac{\sigma}{r_s}\right)^{6} \right] \, H \left( 2^{1/6} \sigma - r_s \right),
	\label{eq:U_LJ}
	\end{equation}
where H is the heaviside step function, $r_s$ the point of closest approach between two components, $\sigma$ the cut-off distance (under which steric interactions occur), and $F_s$ the repulsion strength. Following Ref. \cite{Adhyapak:2015jk}, we set $F_s =$ 0.8 pN and set $\sigma = 4a$ for numerical stability. Our full steric calculations are described in Appendix \ref{app:sterics}, again modeled after Ref. \cite{Adhyapak:2015jk}.

\subsection{Motor}
Following the derivation for propagation of motor torque around a bend presented in Nguyen et al. \cite{Nguyen:2017jt}, we write the motor torque on each flagellum as:
\begin{equation}
\mathbf{T}_1^\mot = \frac{T}{2} \left( \mathbf{e}_{0}^3 + \mathbf{e}_{1}^3  \right) \rightarrow \mathbf{T}_1^\mot \cdot \mathbf{e}_1^3 = \frac{T}{2}(1+\cos\theta_0).
\label{eq:torque_mot}
\end{equation}
Note that the motor torque is only applied to the first segment (hook) of each flagellum. When there is no hook bending, i.e. $\theta_{0}=0$, we simply have $\mathbf{T}_1^\mot \cdot \mathbf{e}_{1}^3 = T$.  For the general case of a bent hook, we decompose $\mathbf{T}^\mot$ to a torque acting along the hook direction $\mathbf{e}_{1}^3$ and forces acting on the adjacent nodes 0 and 1:
\begin{equation}
	\mathbf{F}_{0,1}^\mot = \pm \frac{T}{2L_h} \mathbf{e}_{0}^3 \times \mathbf{e}_{1}^3.
	\label{eq:force_mot}
\end{equation}
%
%
%
By Newton's third law, a corresponding counter-torque $-\mathbf{T}_1^\mot$ is exerted on the body for each flagellum (leading to counter-rotation of the body):
%
\begin{equation}
	\mathbf{T}_b^\mot = \sum_\mathrm{flagella} -\mathbf{T}_1^\mot.
	\label{eq:body_motor}
\end{equation}


\subsection{Numerical methods and simulation}
Hereinafter, we present quantities in nondimensionalized form, scaling torques with $T$, lengths with $R_b$, forces with $T/R_b$, velocities with $T/R_b\zeta_b$ and time with $\zeta_b R_b^2/T$. 
 We choose our standard swimmer geometry (scaled by $R_b$) to be $L_h=0.28$, $R = 0.28$, $\lambda=4.0$, $a=0.028$, $L=9$. \revision{The helix pitch $\lambda=4$ is a good estimate for normal form \textit{E. coli} \cite{Turner:2000vm}.} \revision{The chosen value of $L_h$ is characteristic of a uniflagellar organism.   A larger hook length provides more clearance between the cell body and the flagellum, decreasing the importance of steric interactions and thus the time-step restriction on the numerical simulations. We have performed spot checks on stability boundaries using smaller $L_h$ and found no substantial changes. Each flagellum is discretized into segments (rods) of length $l=0.28$. With this flagellar geometry there are 16 rods per wavelength of flagellum. This level of discretization has been found to be sufficient to yield converged results. }
%

To run the simulation, we initialize the swimmer in its equilibrium state ($\mathcal{E}^\el = 0$) with triads on each edge initialized as in Refs. \cite{Reichert2006,Vogel:2010gc} using Eqs. \ref{eq:e3} and \ref{eq:e1e2}. At $t=0$, torque is applied to each flagellum and we track the resulting motion. Our simulation algorithm is a combination of Fast Projection \cite{Goldenthal2007} and SHAKE-HI \cite{Allison1984}, while adapting the rigid-body coupling formalism of Bergou et al. \cite{Bergou2008}. We use unconstrained dynamics to step forward in time and project until constraints are satisfied. We include the full algorithm details and equations in Appendix \ref{app:numerics}. The projection step accounts for all hydrodynamic interactions, and the constraints are automatically calculated and applied. Using our method, the (dimensionless) time step $\Delta t$ is $\sim$ $10^{-6}$ and we satisfy constraints to within $10^{-12}$. Simulations are run to \revision{28} time units, which is sufficient to allow a stable swimmer to translate several body lengths. As a check on the calculations we note that in the limit of low flexibilities our swimming simulations are comparable to resistive force theory (RFT) calculations -- we do indeed find reasonable agreement and summarize the results in Appendix \ref{app:RFT}. 

\section{Results}



\redtext
\revision{Our presentation of the results is as follows. Sections \ref{sec:uni}-\ref{sec:quad} describe parameter sweeps in $\Fl$ and $\Flh$ that culminate in phase diagrams of different regimes of swimming for bacteria with $N=1,2,3,$ and $4$ flagella anchored at the corners of a regular tetrahedron inscribed in the cell body. To characterize flagellar conformations we use the root mean squared distance $D$ of the flagellar nodes from the center of the body:
	\begin{equation}
	D^2 = \frac{1}{N M} \sum_{\mathrm{nodes}} |\mathbf{x}_i - \mathbf{x}_b|^2,
	\label{eq:Dbar}
	\end{equation}
	where $N M$ is the total number of flagellar nodes. All simulations will start with the flagella sticking out from the body; this is the rest state for the elastic energies. For the standard parameter set given above, $D=0.6455$ at rest.  Accompanying the phase diagrams will be images illustrating the various observed flagellar conformations.  These sections give an overall picture of how the parameter regime for stable swimming changes with flagellar number. Section \ref{sec:robust} presents a brief study of the robustness of quadriflagellar locomotion with bundled flagella with respect to the spatial distribution of anchor points (flagellar motors). Finally, Section \ref{sec:speed} addresses the dependence of swimming speed on flagellar number, with computational results and a simple theoretical result.}

\subsection{Uniflagellar swimming}\label{sec:uni}

\begin{figure}
	\centering
	\begin{minipage}{0.49\linewidth}
		\includegraphics[]{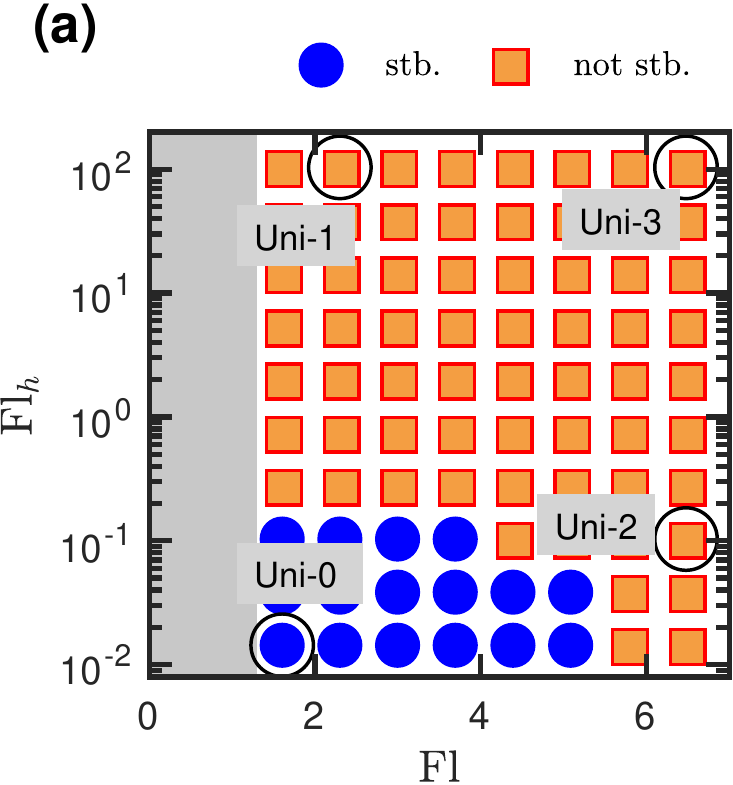}
	\end{minipage}
	\begin{minipage}{0.49\linewidth}
		\includegraphics[]{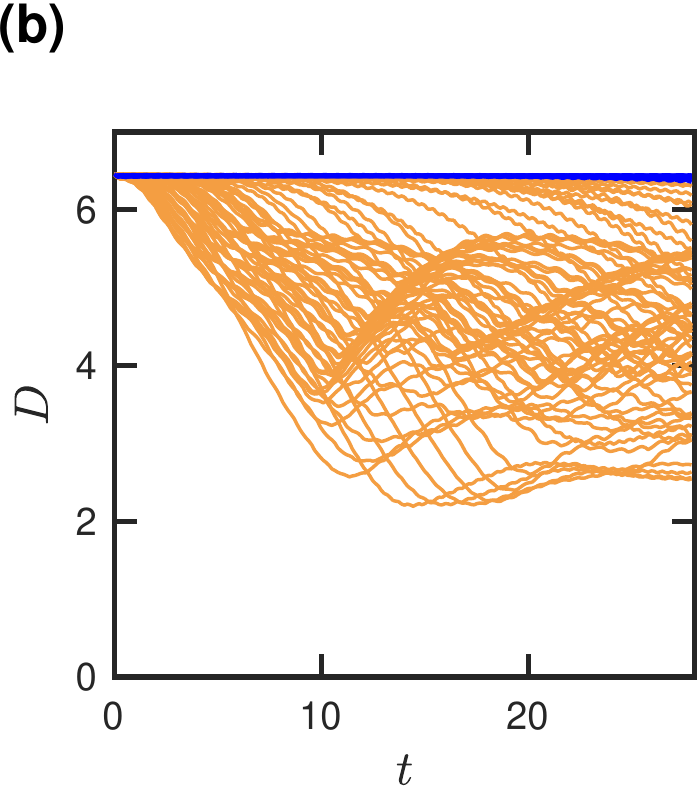}
	\end{minipage}
	\caption{\revision{(a) Phase diagram for uniflagellar swimmer. Blue circles denote stable straight swimming, and orange squares denote conditions where straight swimming is not stable. Black circles denote the cases of ($\Fl$, $\Flh$) marked for study: Uni-0 (1.6, 0.014), Uni-1 (2.3, 100), Uni-2 (6.5,0.1), Uni-3 (6.5,100). (b) Flagellar distance from body $D$ vs. time $t$ color coded as in (a).}}
	\label{fig:bif1n}
\end{figure}

\begin{figure}
	\centering
	\includegraphics[width=\singlecolumnwidth]{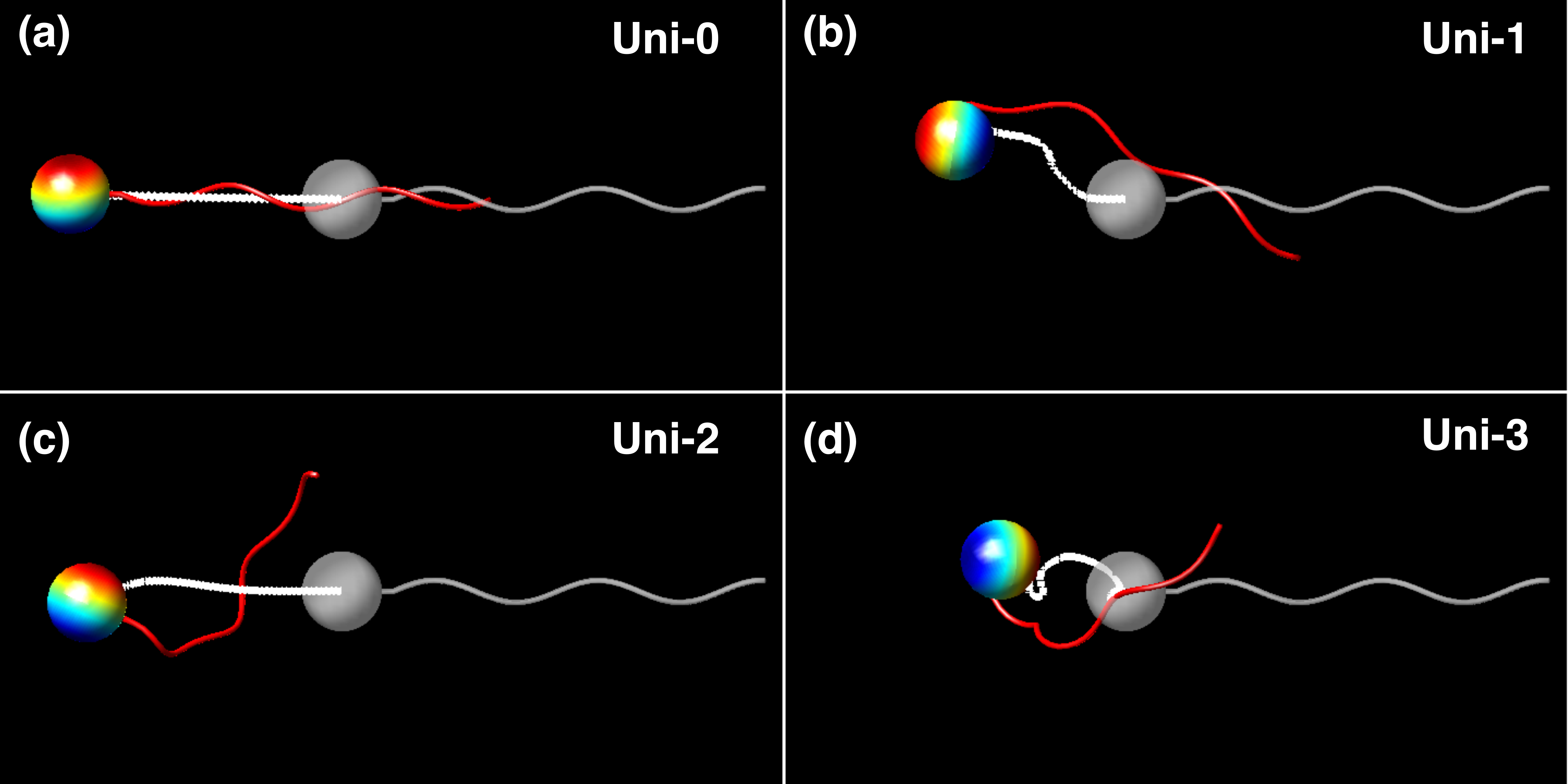}
	\caption{\revision{Snapshots of uniflagellar swimmer at $t=0$ (gray) and $t=28$ (color). The flexibilities ($\Fl$, $\Flh$) for each case are: Uni-0 (1.6, 0.014), Uni-1 (1.6, 100), Uni-2 (6.55,0.1), Uni-3 (6.5,100). Dotted white lines follow the body center.}}
	\label{fig:unisnap_n}
\end{figure}

As described in prior work \cite{Son:2013dh,Nguyen:2017jt}, a uniflagellar bacterium pushed by its flagellum is susceptible to buckling of the hook protein. Here we generalize that result, presenting in Fig. \ref{fig:bif1n}(a) a phase diagram for uniflagellar swimming. Note that the scales for $\Fl$ and $Fl_h$ are linear and logarithmic, respectively. We classify the trajectories into two categories. Blue circles represent stable swimming in an essentially straight path (actually there is a slight helical nature to the trajectory, as described in \cite{Nguyen:2017jt}). Orange squares indicate cases where straight swimming is not stable. For the simple model in \cite{Nguyen:2017jt}, this transition actually arises via a supercritical Hopf bifurcation from a steady and very small angle between body and flagellum to a time-periodic angle whose average value increases as the flexibility number increases above the critical value. In Fig.~\ref{fig:bif1n}(b), we show the corresponding time evolution of the distance metric $D$ for all the parameter values in Fig.~\ref{fig:bif1n}(a) and color-coded accordingly. For the straight-swimming cases, $D$ remains near the equilibrium value $D=0.6455$, indicating that the flagellum remains nearly straight and extended essentially normal to the body. In all other cases, $D$ deviates downward, indicating distortions of the hook, flagellum or both. 

We now present further details at four points on the phase diagram.
The first case, labeled ``Uni-0' in Figure \ref{fig:bif1n}, is for  $\Fl=1.6,\Fl_h=0.014$. The motion of this swimmer, which we describe as ``stable straight swimming" is depicted by the snapshot in Figure \ref{fig:unisnap_n}(a). The gray image is the cell at time $t=0$ and the color one is at $t=28$. A white line shows the trajectory of the body center position $\mathbf{x}_b$. Here the flagellum rotates clockwise with negligible deformation to produce constant thrust on the body, which slowly rotates counter-clockwise as it translates. The swimming trajectory is virtually linear with speed $v_b\approx 0.24$, and the hook is virtually undeformed. We note that at $t=0$, the flagellar axis is not aligned with the center of a cell because the flagellum is a pure helix. However, once the flagellum begins rotating, the flagellar shape tapers slightly near the body and the flagellar helical axis aligns with the center of the body during flow. Despite this alignment at steady-state, the hook angle $\theta_0$ is not exactly zero (though still very small) due to the generation of a slight thrust normal to the helical axis as described in \cite{Nguyen:2017jt} for finite-length flagella. As we noted above and in Fig. \ref{fig:bif1n}(a), straight uniflagellar swimming is stable only in a small region of flexibility parameter space, i.e. only at the lowest flexibilities, and so now we describe swimming behavior away from the stability boundary to understand how locomotion changes once instability occurs.

Fig. \ref{fig:unisnap_n}(b) illustrates case ``Uni-1," where $\Fl = 2.3$ and $\Flh = 100$. We have increased $\Fl_h$ by several orders of magnitude from Uni-0 while maintaining a stiff flagellum. In this snapshot, we see that the swimmer enters a slow and broad helical wobble with the flagellum extending nearly tangent to the body -- the cell has buckled at the hook.
This phenomenon has an experimental analogue; as Son et al. showed that a \textit{V. alginolyticus} cell with a buckled hook also exhibits a curved trajectory \cite{Son:2013dh}. Those authors point out that this instability enables a uniflagellar cell to reorient and thus better explore its environment. 

We now consider the case ``Uni-2" with $\Fl = 6.5$ and $\Flh = 0.1$. In contrast to Uni-0 and Uni-1, here we have a swimmer with a rather stiff hook and very flexible flagellum. From the Fig. \ref{fig:unisnap_n}(c) snapshot, we see that the swimmer maintains a straight path for about 15 time units (about 3 times longer than Uni-1) before deviating from the Uni-0 trajectory. Though it is not apparent from the snapshot, the trajectory does eventually stall at long time. Here the flagellar filament is buckling, but in contrast to the hook buckling case Uni-1,  the cell does not reorient.  
The distorted flagellar helix observed here is similar to the those reported in \cite{Jawed:2015vw,Vogel:2012bm}. This particular buckling behavior is, to our knowledge, not seen in nature, perhaps because unlike hook buckling, pure flagellar buckling impedes swimming without the auxiliary benefit of trajectory reorientation.

Finally, for completeness, we present case ``Uni-3" with ($\Fl$, $\Flh$) = (6.5,100) in Fig. \ref{fig:unisnap_n}(d) -- here substantial bending of both the hook and flagellum are observed, and the cell cannot successfully swim any distance.
%
%

\subsection{Biflagellar swimming}

\begin{figure}
	\centering
	\begin{minipage}{0.49\linewidth}
		\includegraphics[]{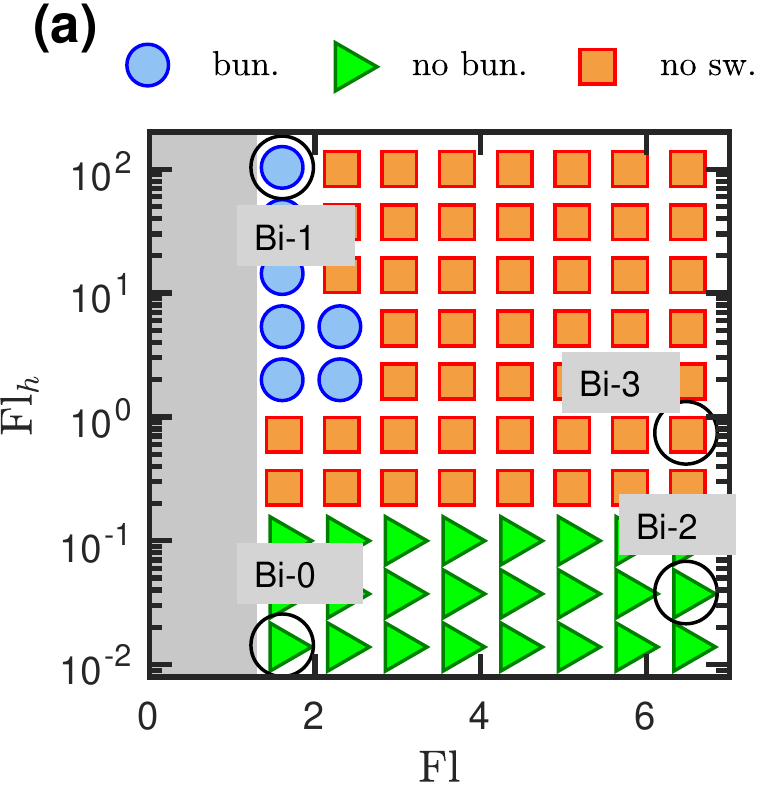}
	\end{minipage}
	\begin{minipage}{0.49\linewidth}
		\includegraphics[]{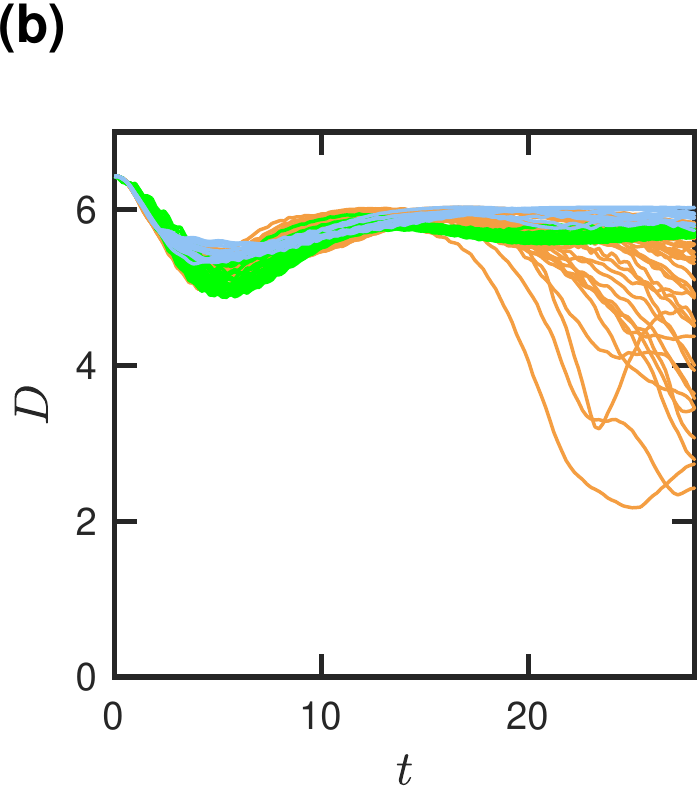}
	\end{minipage}
	\caption{\revision{(a) Phase diagram for biflagellar swimmer with $\lambda=4$. Blue circles denote swimming with an intermittent bundle, green triangles denote swimming with no bundling and orange squares denote ineffective swimming. Black circles denote the cases of ($\Fl$, $\Flh$) marked for study: Bi-0 (1.6, 0.014), Bi-1 (1.6, 100), Bi-2 (6.5,0.038), Bi-3 (6.5,0.72). (b) Flagellar distance from body $D$ vs. time $t$. }}
	\label{fig:bif2n}
\end{figure}

\begin{figure}
	\centering
	\includegraphics[width=\singlecolumnwidth]{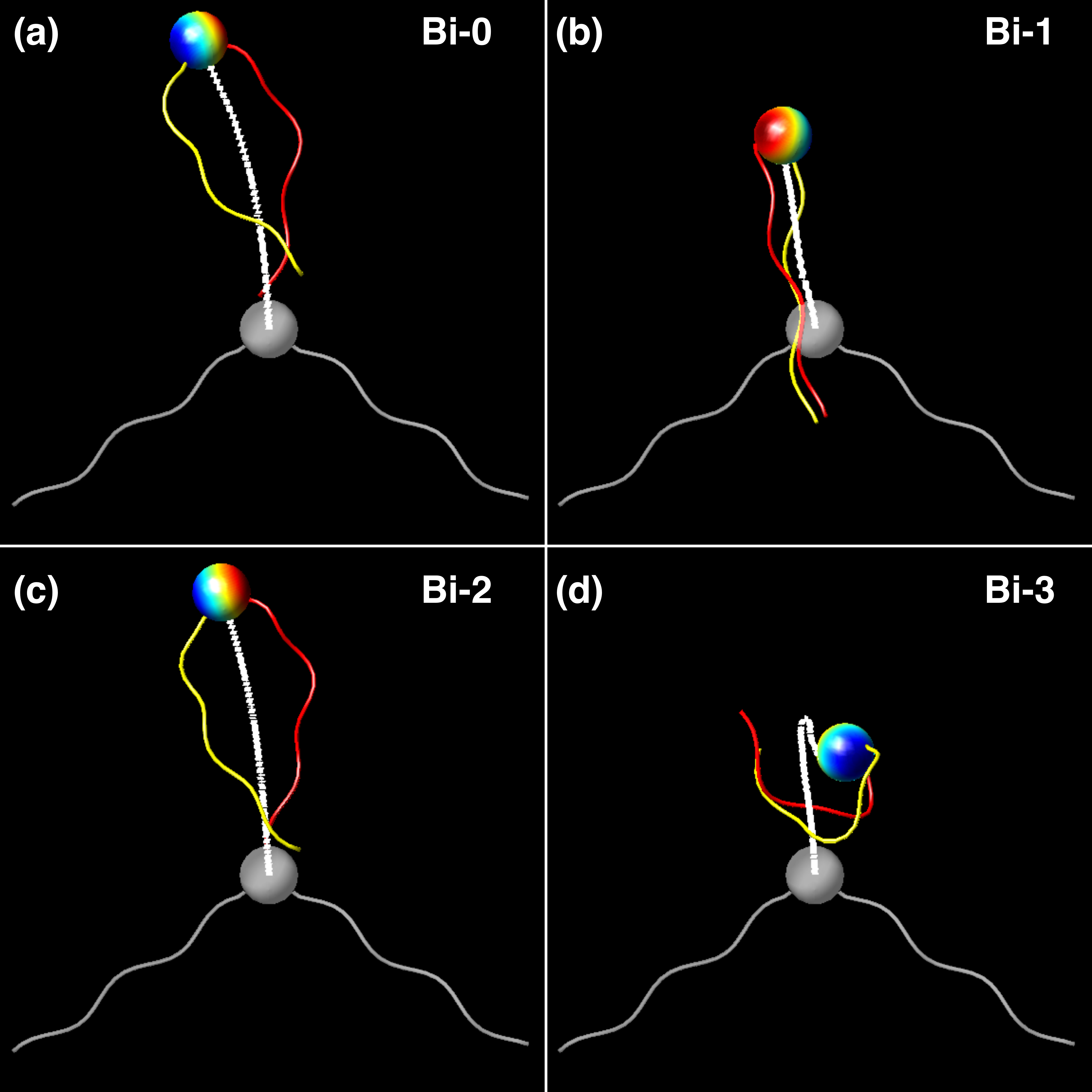}
	\caption{\revision{Snapshots of quadriflagellar swimmer with flagellar anchors placed at the vertices of a regular tetrahedron at $t=0$ (gray) and $t=28$ (color). The flexibilities ($\Fl$, $\Flh$) for each case are: Bi-0 (1.6, 0.014), Bi-1 (1.6, 100), Bi-2 (6.5,0.038), Bi-3 (6.5,0.72). Dotted white lines follow the body center.}}
	\label{fig:bisnap_n}
\end{figure}

The phase diagram for biflagellar swimming is shown in Fig.~\ref{fig:bif2n}. Observe that we now have three categories. Blue symbols represent exhibit swimming with the formation of  bundles, while green symbols represent swimming without bundling and orange symbols indicate the absence of coherent swimming. 
Viewed in terms of $D(t)$, the blue and green cases split into two fairly distinct bands, as illustrated in Fig.~\ref{fig:bif2n}(b). In general, the cases that swim with bundles (blue) have larger $D$ than those that swim without (green), indicating that the bundled flagella extend have a straighter conformation than the unbundled ones. Cells that do not swim generally have curved and disorganized flagella with relatively small $D$. While we will not pursue this issue here, the distinction between bundled and unbundled states is fairly sharp, as can be seen by the plots of $D(t)$, suggesting that these regimes are separated by bifurcations. Janssen and Graham \cite{Janssen:2011ex} have explored bifurcations between bundled and unbundled states for pairs of anchored flagella. 


Swimming without bundles occurs when the hook is stiff, as in Bi-0, where ($\Fl$, $\Flh$) = (1.6,0.014) and Bi-2, where ($\Fl$, $\Flh$) = (6.5,0.038).  Observe that the regime of stable biflagellar swimming without bundles is similar to the regime of uniflagellar stable swimming, although it extends to higher values of $\Fl$. Snapshots of these two cases are shown in Figs.~\ref{fig:bisnap_n}(a) and (c). In each case, the swimmer translates in a slightly curved path. Though the flagella are in closer proximity than their starting positions (in gray), they clearly do not bundle, remaining remain splayed out as the stiff hooks do not permit the flagella bend very much toward one another. Further away from the body, Bi-2 shows slightly more deformation down its length than Bi-0 (by about 1 helical turn) due to the higher $\Fl$. 
The speed in both cases is faster than Uni-0 by about 50\%.

Bundling can only occur if the hooks are sufficiently flexible to allow the flagella to come together and the flagella are sufficiently stiff not to buckle. We examine case Bi-1, where ($\Fl$, $\Flh$) = (1.6,100).  Fig. \ref{fig:bisnap_n}(b) shows a snapshot that indicates formation of a loose bundle.
The Bi-1 swimmer translates with almost an identical speed to Uni-0, but we note that it is stably swimming at an $\Flh$ about 4 orders of magnitude larger than Uni-0, and of course this regime of stable swimming is completely absent from the uniflagellar phase diagram. In combination with the extension of the region of stable swimming at low $Fl$, this result indicates that the change from uni- to biflagellar swimming confers substantial stability for straight swimming. 

We point out that although we see more stable swimming overall for the biflagellar swimmer than the uniflagellar case, ineffective swimming (``flailing" would be a good description) still occupies a significant portion of parameter space. We show one example of a poor swimmer with case Bi-3, ($\Fl$, $\Flh$) = (6.5,0.72), whose snapshot is depicted in Fig. \ref{fig:bisnap_n}(d). While the swimmer initially appears to swim, it doubles back on its trajectory, and the flagella ultimately end up contorted around the cell body and the body trajectory stalls.


\subsection{Triflagellar swimming}

\begin{figure}
	\centering
	\begin{minipage}{0.49\linewidth}
		\includegraphics[]{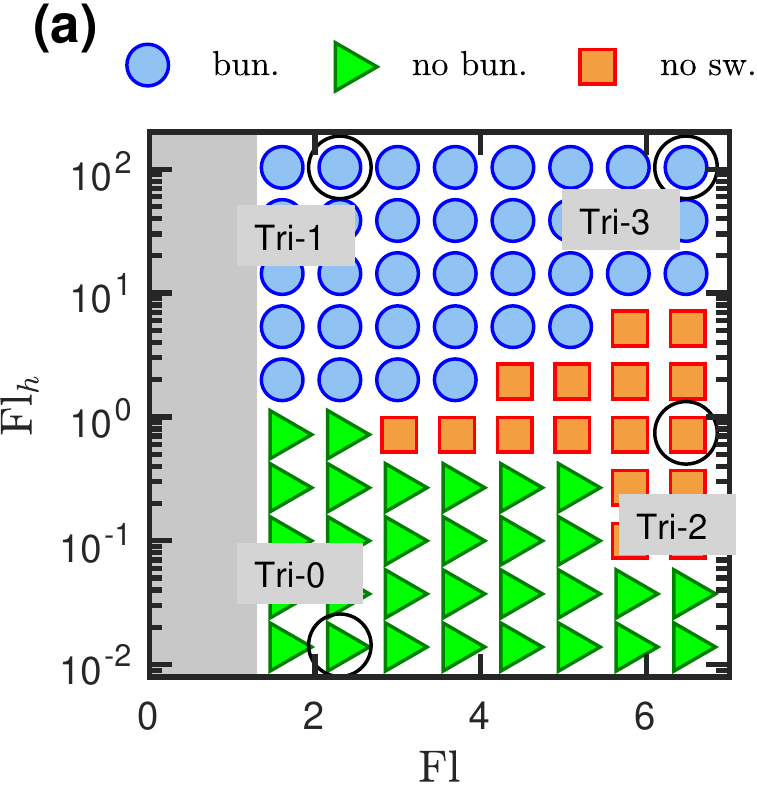}
	\end{minipage}
	\begin{minipage}{0.49\linewidth}
		\includegraphics[]{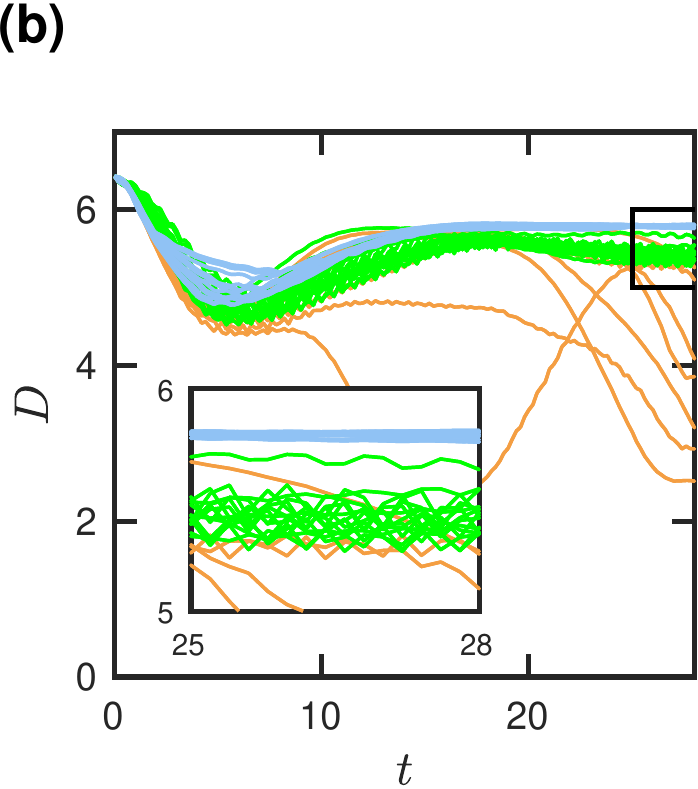}
	\end{minipage}
	\caption{\revision{(a) Phase diagram for triflagellar swimmer. Blue circles denote swimming with a stable bundle, green triangles denote swimming with no bundling and orange squares denote ineffective swimming. Black circles denote the cases of ($\Fl$, $\Flh$) marked for study: Tri-0 (2.3, 0.014), Tri-1 (2.3, 100), Tri-2 (6.5,0.72), Tri-3 (6.5,100). (b) Flagellar distance from body $D$ vs. time $t$.  The inset shows the demarcation in $D$ for the bundling (blue) and non-bundling (green).}}
	\label{fig:bif3n}
\end{figure}

\begin{figure}
	\centering
	\includegraphics[width=\singlecolumnwidth]{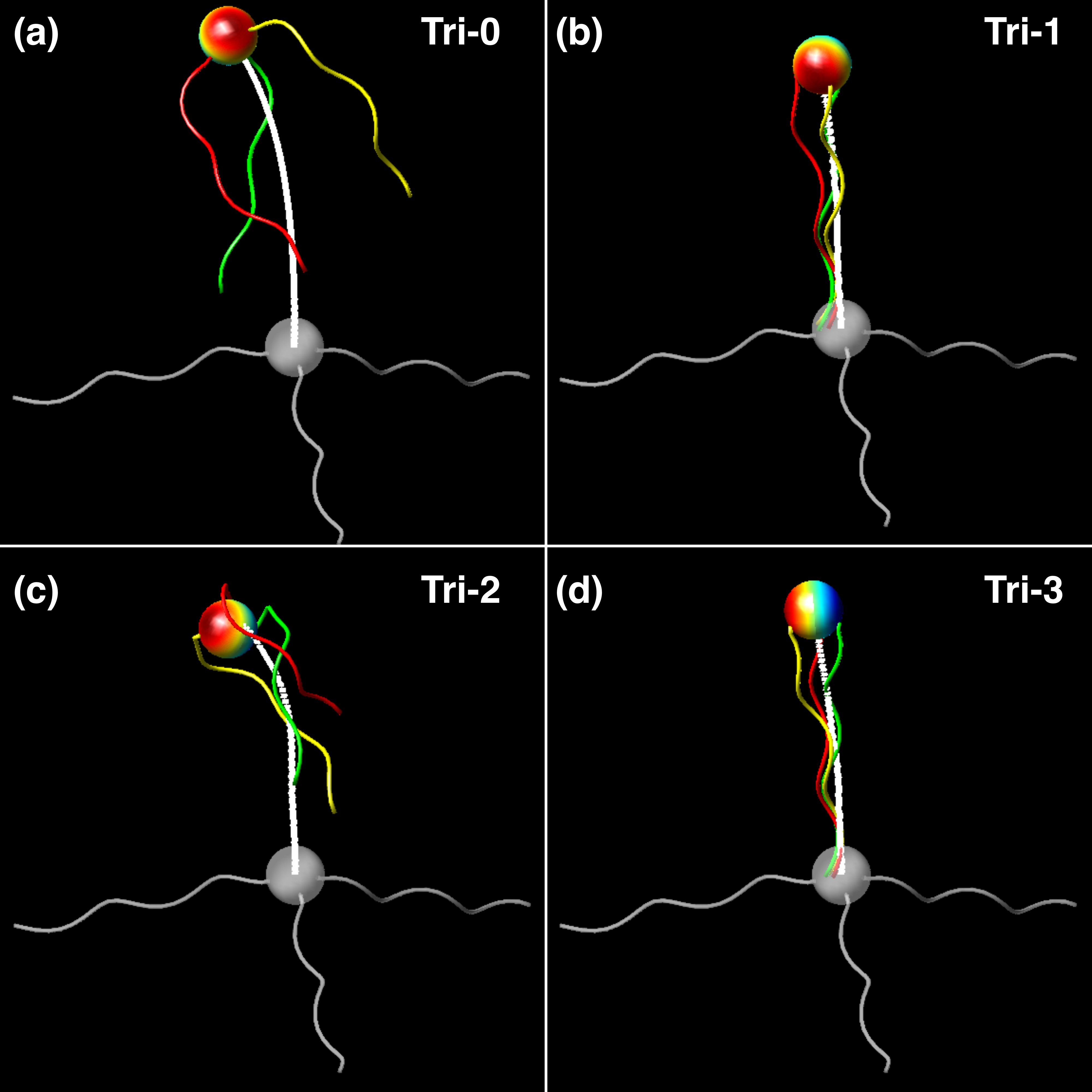}
	\caption{\revision{Snapshots of triflagellar swimmer at $t=0$ (gray) and $t=28$ (color). The flexibilities ($\Fl$, $\Flh$) for each case are: Tri-0 (2.3, 0.014), Tri-1 (2.3, 100), Tri-2 (6.5,0.72), Tri-3 (6.5,100). Dotted white lines follow the body center $\mathbf{x}_b$ over 28 time units.}}
	\label{fig:trisnap_n}
\end{figure}

Further dramatic changes in the stability of straight swimming arise, when going from bi- to triflagellar cells. Fig.~\ref{fig:bif3n}(a) shows the triflagellar phase diagram, showing greatly expanded regions of stable bundled or unbundled swimming.
This categorization is again reflected in the bands observed in $D(t)$ (Fig. \ref{fig:bif3n}(b)), with the distinction being sharper in the triflagellar case than the biflagellar. As shown in the inset of Fig. \ref{fig:bif3n}(b), the blue lines, again reflecting bundled swimming, collapse almost uniformly near $D=5.8$, and the green lines, reflecting swimming without bundles,  show a lower time-average, are larger deviations.  

Turning to specific cases, we first look at swimming without bundling,  case ``Tri-0" (Fig. \ref{fig:trisnap_n}(a)), where ($\Fl$, $\Flh$) = (2.3,0.014). Analogous to the biflagellar case, the cell here swims in a slightly curved path with its flagella splayed in a tripod-like configuration. The swimming speed here is about 60\% higher than Uni-0, and not much faster than Bi-0 or Bi-2. Similar results are obtained in the green region of Fig.~ \ref{fig:bif3n}(a) even when increasing $\Fl$. Thus at low $\Flh$, the three flagella do not seem to confer significantly more stability or speed than two flagella.

At higher $\Flh$, however, we see substantial changes from the biflagellar case. Above a value of $\Flh\sim 1$, stable swimming with bundled flagella is found over a wide range of $\Fl$, a marked expansion from the biflagellar case  (Fig. \ref{fig:bif2n}(a)). We show two examples of this stable region:  cases ``Tri-1", where ($\Fl$, $\Flh$) = (2.3,100) and ``Tri-3", where ($\Fl$, $\Flh$) = (5,100). Corresponding snapshots are shown in Figs. \ref{fig:trisnap_n}(b) and (d). Here the flagella gather behind the cell from their starting positions and bundle to push the cell forward. This is the classic behavior seen for peritrichous bacteria like \textit{E. coli}, as noted in \cite{Berg2003, Darnton:2007ct}. 
The swimming speed in these cases is about 40\% higher than Uni-0, and we emphasize that $\Flh$ for the cases here are several orders of magnitude higher than the stable swimming uniflagellar swimming case Uni-0.

For completeness, we also show one case, ``Tri-2", where ($\Fl$, $\Flh$) = (6.5,0.72), that does not swim effectively (Fig. \ref{fig:trisnap_n}(c)). 
Overall, however, we see quite a remarkable stability transformation from two to three flagella, particularly at high $\Fl$ values associated with peritrichous bacteria. 

\subsection{Quadriflagellar swimming}\label{sec:quad}

\begin{figure}
	\centering
	\begin{minipage}{0.49\linewidth}
		\includegraphics[]{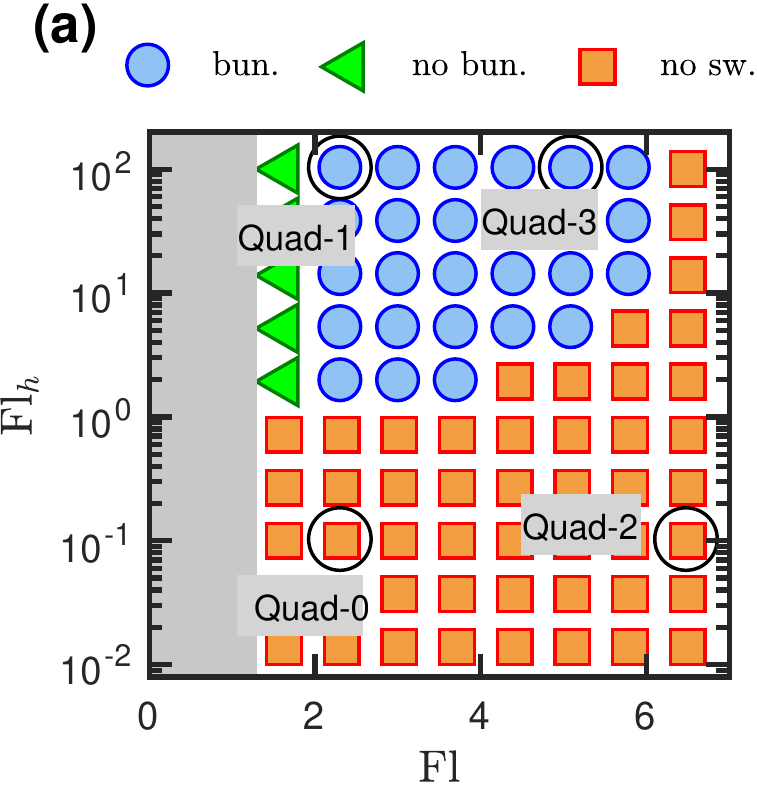}
	\end{minipage}
	\begin{minipage}{0.49\linewidth}
		\includegraphics[]{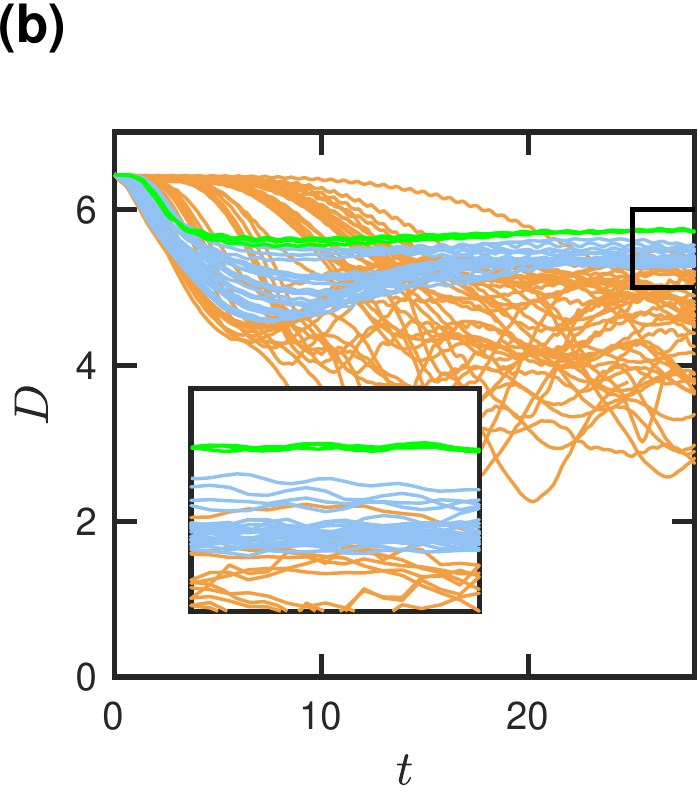}
	\end{minipage}
	\caption{\revision{(a) Phase diagram for quadriflagellar swimmer. Blue circles denote swimming with a stable bundle, green triangles denote no bundling, and orange squares denote ineffective swimming. Black circles denote the cases of ($\Fl$, $\Flh$) marked for study: Quad-0 (2.3, 0.1), Quad-1 (2.3, 100), Quad-2 (6.5,0.1), Quad-3 (5.1,100). (b) Flagellar distance from body $D$ vs. time $t$.  The inset shows the demarcation in $D$ for the bundling (blue) and non-bundling (green).}}
	\label{fig:bif4n}
\end{figure}

\begin{figure}
	\centering
	\includegraphics[width=\singlecolumnwidth]{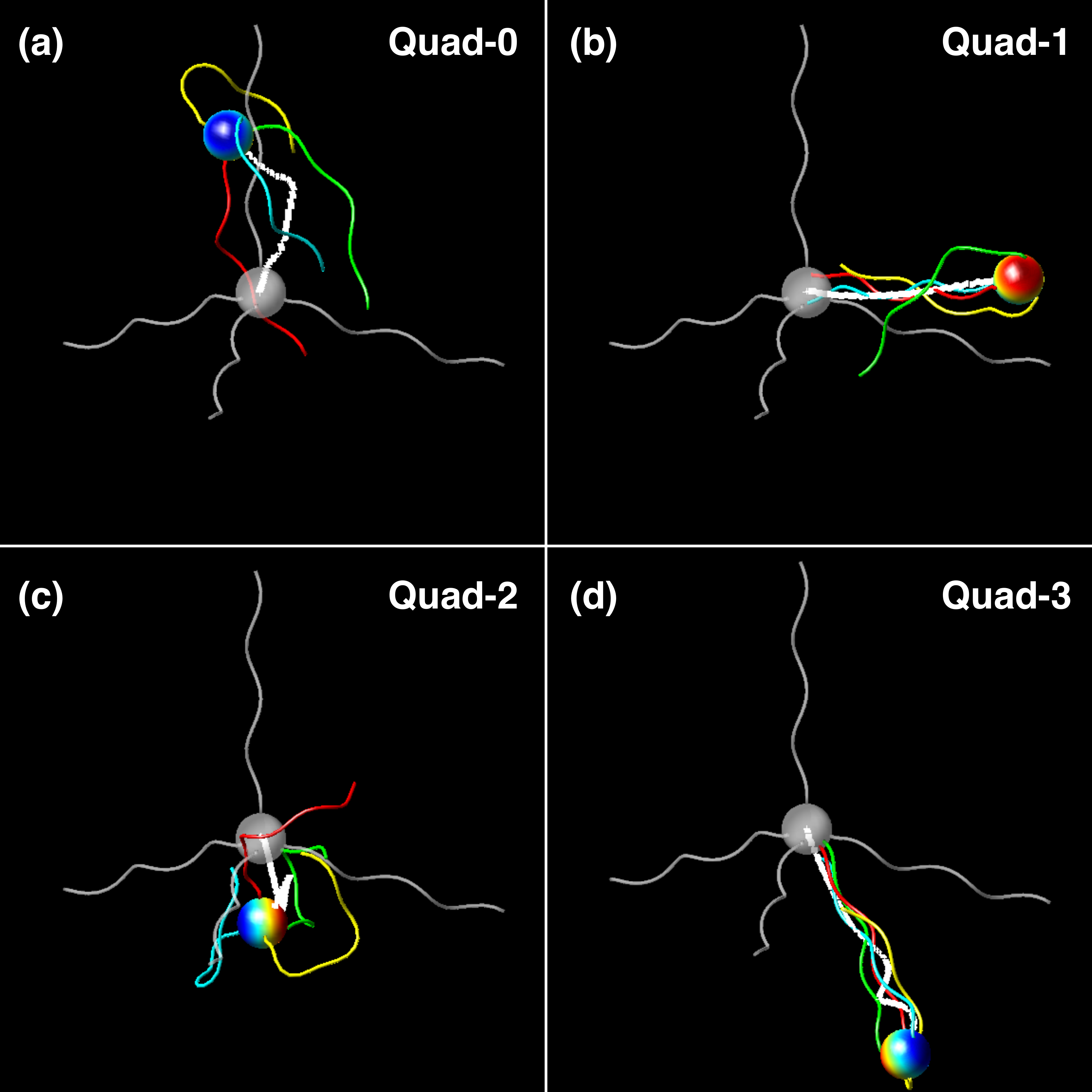}
	\caption{\revision{Snapshots of quadriflagellar swimmer at $t=0$ (gray) and $t=28$ (color). The flexibilities ($\Fl$, $\Flh$) for each case are: Quad-0 (2.3, 0.10), Quad-1 (2.3, 100), Quad-2 (6.5,0.1), Quad-3 (5.1,100). Dotted white lines follow the body center $\mathbf{x}_b$ over 28 time units.}}
	\label{fig:tetrasnap_n}
\end{figure}

A dramatic change in the phase diagram again occurs upon the addition of a fourth flagellum, so that flagella are anchored at all four corners of the inscribed tetrahedron. The phase diagram is shown in Fig.~\ref{fig:bif4n}(a).
Here $D(t)$ (Fig.~\ref{fig:bif4n}(b)) shows the same general banding as the bi- and triflagellar cases, so we again use the same labels to classify swimming. One difference here compared to the triflagellar case is that $D$ for the stable bundled cases (blue) can be have substantial fluctuations while the unbundled swimming cases (green) are quite smooth.
Like the triflagellar case, the quadriflagellar cell swims quite stably at high $\Flh$, with the same approximate threshold of $\Flh\sim 1$. What we do not see here are many cases of swimming with no bundle, and in particular when $\Flh\lesssim 1$ stable swimming does not occur, with our without bundling. Thus for this quadriflagellar cell, swimming \emph{requires} high hook flexibility. Indeed the regimes of stable swimming for uni- and quadriflagellar swimmer are virtually converses of one another -- stable swimming in one case is unstable in the other.

To elaborate on quadriflagellar swimming dynamics, we first examine case ``Quad-0", where ($\Fl$, $\Flh$) = (2.3, 0.10), the stiff parameter set yielding stable swimming for Uni-0. From the trajectory snapshot in Fig. \ref{fig:tetrasnap_n}(a), we see that 
the body trajectory is not straight at all, and furthermore, we see no evidence of bundling. While in the bi- and triflagellar cases the splayed flagella can push the cell body in one direction, leading to swimming without bundling, this is not the case here due to the isotropic flagellar arrangement. While body rotation may act to bring the flagella closer together, promoting bundling \cite{Powers:2002p4790}, the very stiff hooks prevent any of the flagella from wrapping around the body to approach one another. The stiff filaments further inhibit any flagellar deformations favorable to bundle formation. Without a bundle, the flagella do not generate thrust in a consistent direction. We note that even at higher $\Fl$, as in case ``Quad-2", where ($\Fl$, $\Flh$) = (6.5,0.1), increased flagellar deformation will not yield bundles without sufficient hook compliance, as illustrated in Fig. \ref{fig:tetrasnap_n}(c). These observations represent a stark change from the uniflagellar case (i.e.~Uni-0), where stiff hooks and filaments are necessary for steady swimming.

We next examine case ``Quad-1" where ($\Fl$, $\Flh$) = (2.3, 100), where the swimmer has stiff flagella and very weak hooks. Snapshots for this swimmer are shown in Fig. \ref{fig:tetrasnap_n}(b). As the body begins rotating, the hooks will bend, most to angles of nearly $\pi/2$, and allow sections of the flagella near the body to wrap around it. Hence, we see that the flagella will also slightly deform near the body as we saw in Uni-1. The flagella proceed to gather on the same side of the body and eventually form a loose bundle that propels the body in a straight path, as shown by the end snapshot of Fig. \ref{fig:tetrasnap_n}(b). By ``loose bundle'', we mean that although the flagella come into contact and move coherently, they do not mesh together at the end to become nearly parallel. 

Lastly, we examine case ``Quad-3", where ($\Fl$, $\Flh$) = (5.1,100); both hooks and filaments are flexible.
Fig. \ref{fig:tetrasnap_n}(d) shows that for this case stable swimming is achieved, with a distinctly helical trajectory. In contrast to Quad-1, with these flexibility parameters, the hooks readily bend and the flagella easily wrap around the cell body and assemble into a tight bundle. 


\begin{figure}
	\centering
	\includegraphics[width=\singlecolumnwidth]{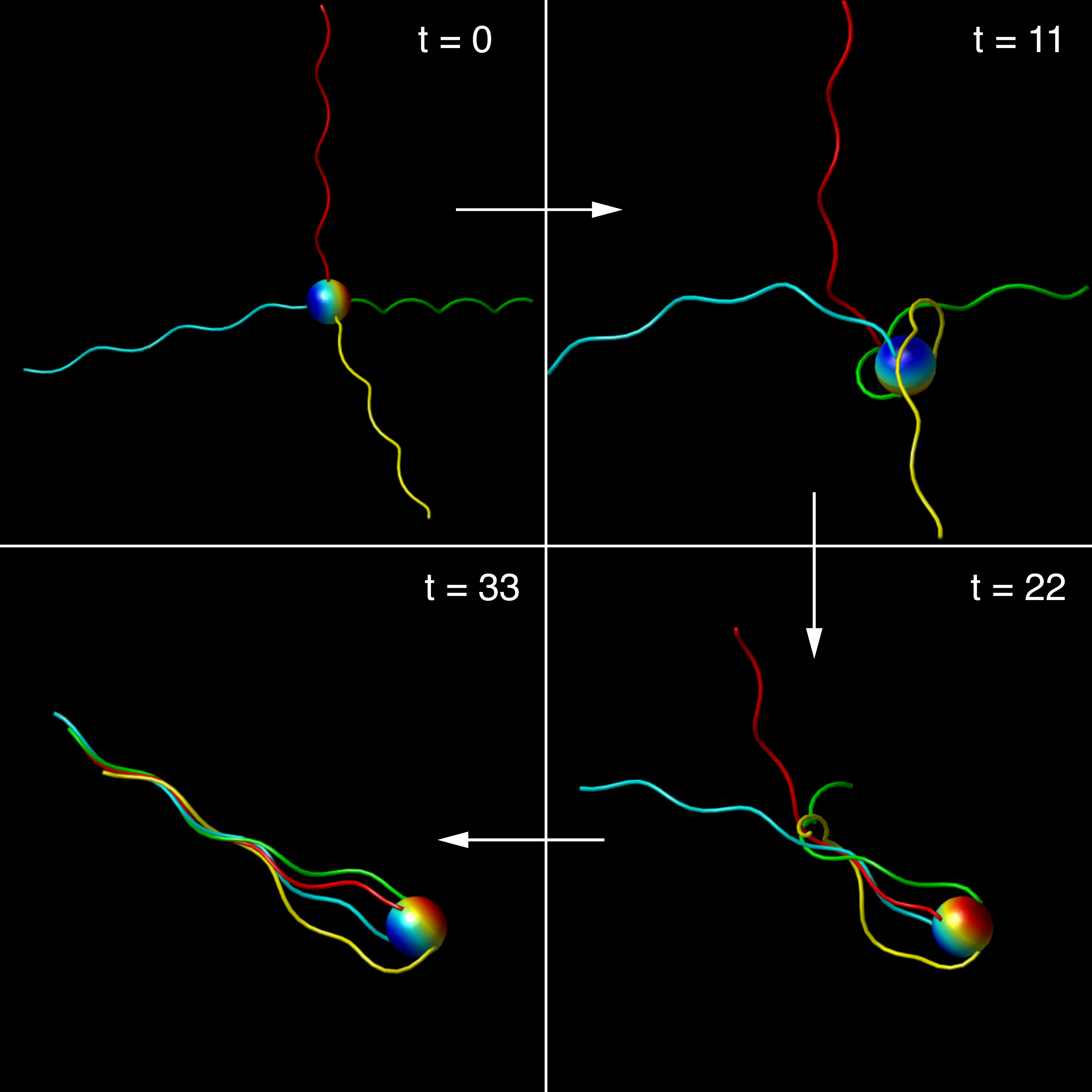}
	\caption{\revision{Sequence of snapshots of the bundling for a quadriflagellar swimmer with longer flagella than in the previous figure ($L=12$). Note that the panel for $t=0$ is zoomed out to show the full swimmer in its original isotropic configuration. Flexibilities are ($\Fl$, $\Flh$) = (5.1,100).}}
	\label{fig:tetra_bundle_tight}
\end{figure}

A more dramatic example of tight bundling is shown in Fig.~\ref{fig:tetra_bundle_tight}, where the flexibility parameters  are ($\Fl$, $\Flh$) = (5.1,100), as for Quad-3, but the flagella have length $L=12$ rather than $8$. The time progression of the bundling process is shown here; the overall process in Fig . \ref{fig:tetra_bundle_tight} has been described as ``zipping and entanglement" by Adhyapak et al. \cite{Adhyapak:2015jk}.


\subsection{Robustness with respect to flagellar arrangement on body surface}\label{sec:robust}

\begin{figure}
	\begin{minipage}{0.37\linewidth}
		\includegraphics[width=\linewidth]{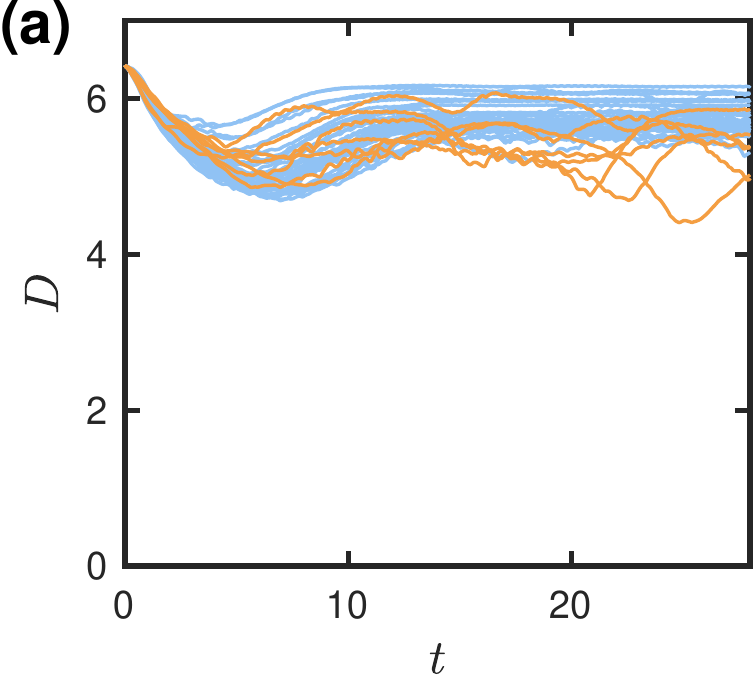}
	\end{minipage}
	\begin{minipage}{0.28\linewidth}
		\vspace{-8mm}
		\includegraphics[width=\linewidth]{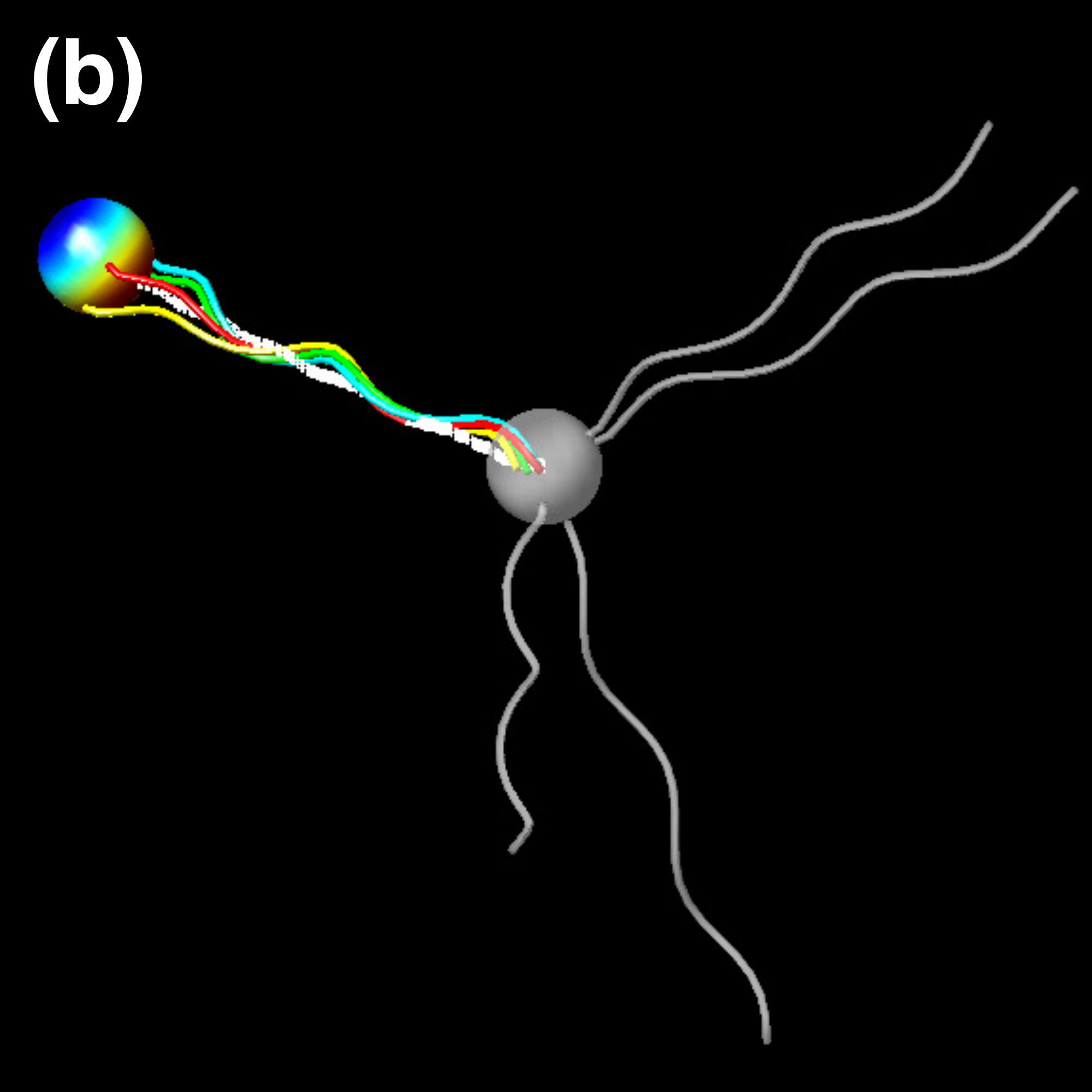}
	\end{minipage}
	\begin{minipage}{0.28\linewidth}
		\vspace{-8mm}
		\includegraphics[width=\linewidth]{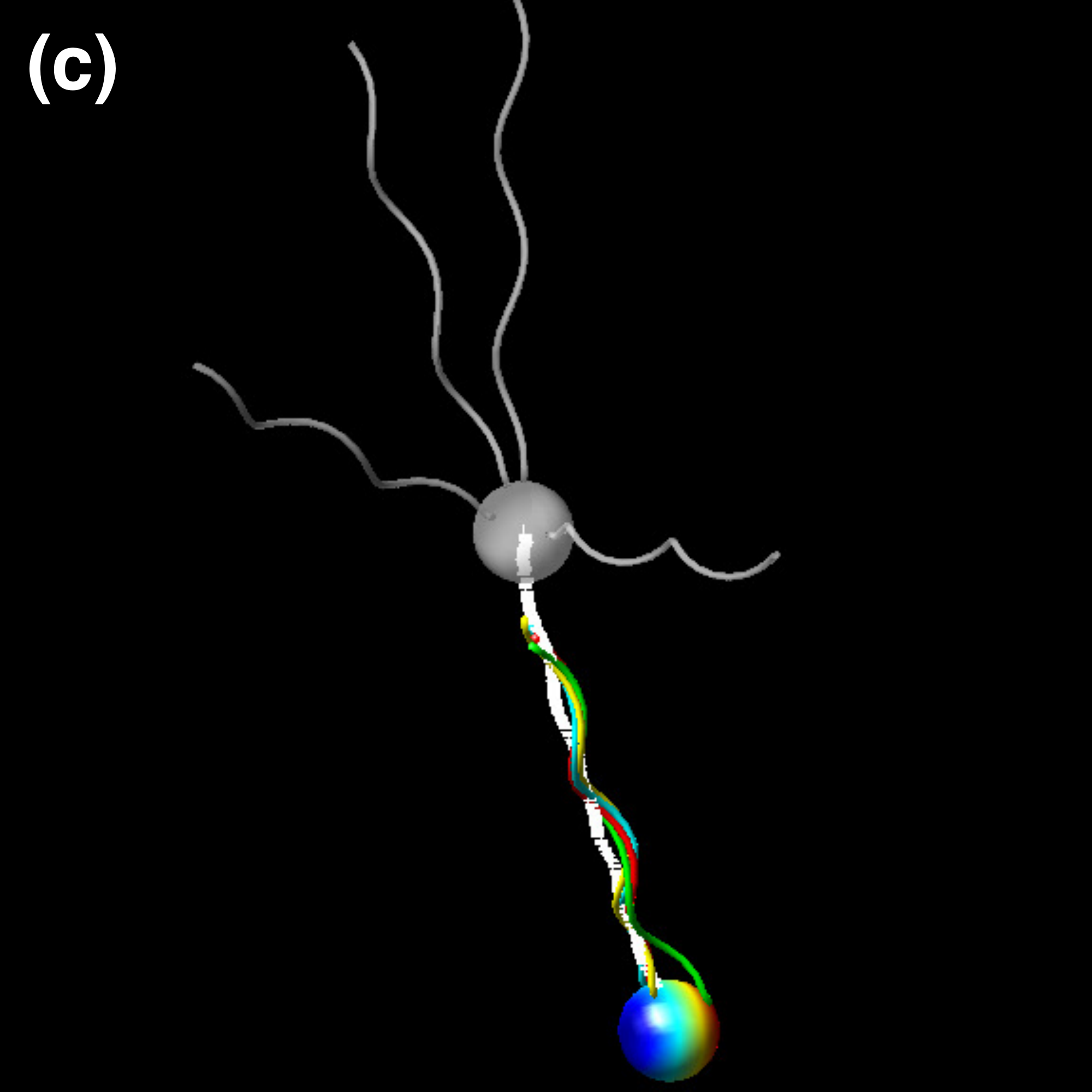}
	\end{minipage}
	\caption{\revision{(a) Plots of $D$ vs. $t$  for swimmers with four randomly sampled flagella with ($\Fl$, $\Flh$) = (4,100). (b) and (c) Examples of stable swimming with randomly anchored flagella: $t=0$ (gray) and $t=28$ (color).}}
	\label{fig:tetra_sample}
\end{figure}

To test the robustness of straight swimming against flagellar arrangement, we consider the parameter set Quad-3 ($\Fl$, $\Flh$) = (4,100), but rather than arranging the flagellar motors tetrahedrally, we generate a large number of samples with flagella located at four randomly chosen locations on the cell surface. Fig.~\ref{fig:tetra_sample}(a) shows $D(t)$ for 40 different samples. For these parameters, 83\% of the samples are able to form bundles and swim, indicating that the ability to swim is fairly robust. Figs~\ref{fig:tetra_sample}(b) and (c) show examples of successful swimming. 



\subsection{Multiflagellarity and swimming speed}\label{sec:speed}

\begin{figure}
	\centering
	\begin{minipage}{0.42\linewidth}
		\centering{\includegraphics[width=\linewidth]{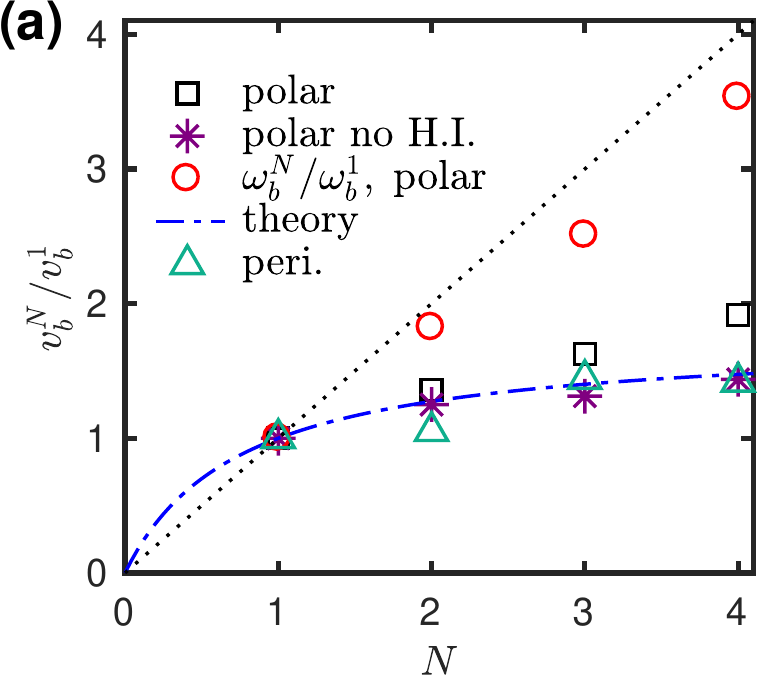}}
	\end{minipage}
	\hspace{5mm}
	\begin{minipage}{3in}
		\centering{\includegraphics[width=\linewidth]{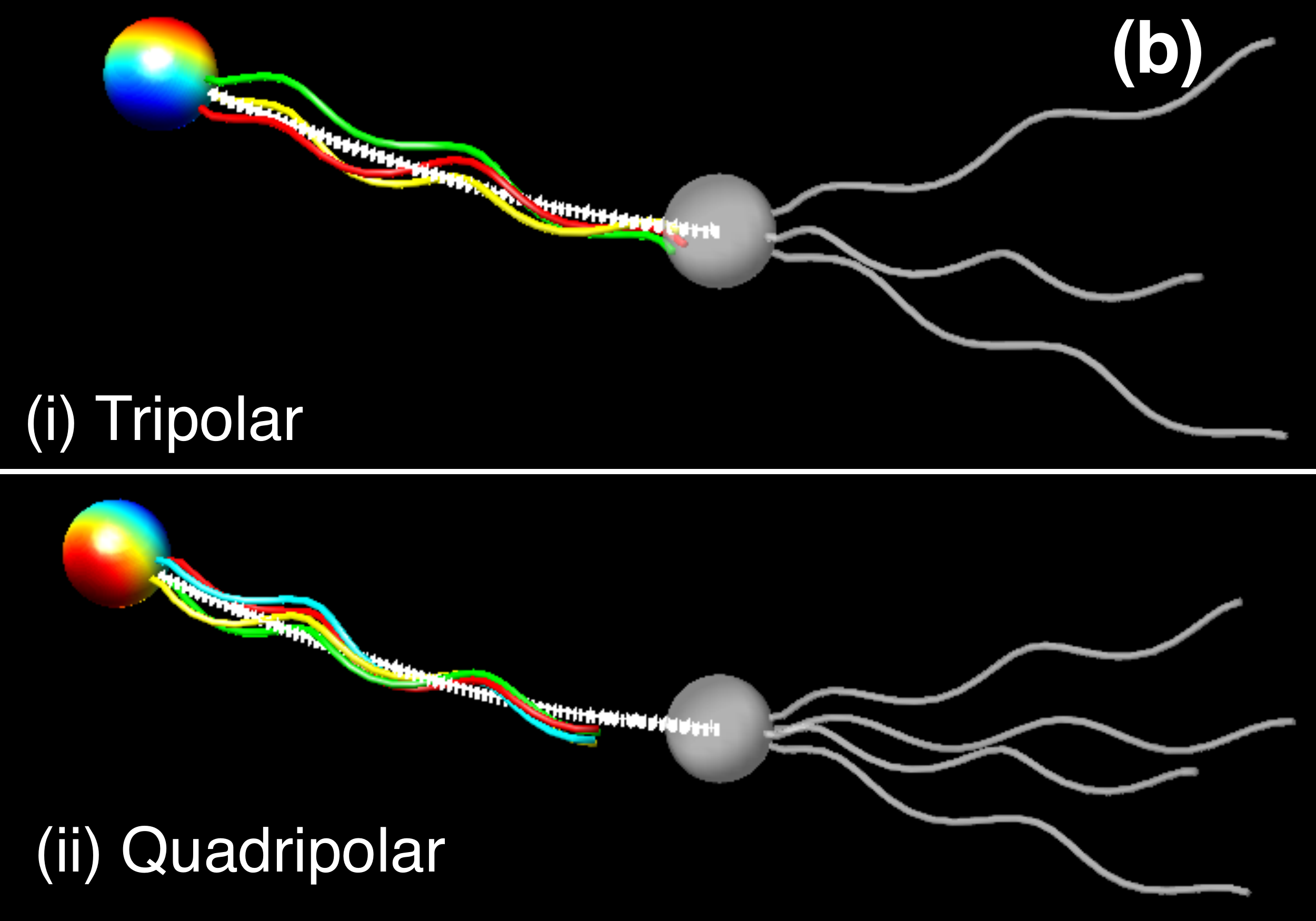}}
	\end{minipage}
	\caption{\revision{(a) Speed of body (normalized to uniflagellar value) vs.~number of flagella. For cells with polar flagella ($\Fl=2.5,\Flh=0.1$), squares are translational speed, $v_b^N/v_b^1$, and circles rotational speed, $\omega_b^N/\omega_b^1$. Asterisks are $v_b^N/v_b^1$ calculated with  no HI. Triangles are $v_b^N/v_b^1$ for the cases Uni-0, Bi-1, Tri-3 and Quad-3. The dashed black line is a linear trend and the solid blue line is Eq. \ref{eq:speed_multiplier}.  (b) Snapshots of swimming for a polar arrangement of (i) three flagella and (ii) four flagella with $t=0$ in gray and $t=21$ in color.}}
	\label{fig:multiflag}
\end{figure}

In the final part of this section, we look at the effects of multiflagellarity on swimming speed. Experimental observations \cite{Darnton:2007ct} indicate that swimming speed does not increase linearly with the number of flagella. We present here simulation and theoretical results that corroborate and explain this observation.

Figure \ref{fig:multiflag} shows results for swimming speed normalized with the corresponding uniflagellar result, plotted against the number of flagella. Two set of computations are shown. The first, labeled ``peri.'', come from the results above, cases Uni-0, Bi-1, Tri-3 and Quad-3, all of which swim stably (with bundles in the multiflagellar cases). The swimming speed seems to saturate by $N=4$ at a speed only about 50\% larger than the unflagellar speed. The other set of computational results, labeled ``polar" comes from simulations of model lophotrichous swimmers with closely spaced polar flagella. That is, we arrange equally-spaced flagellar motors around a small circle on the body surface with solid angle $\varphi_\mathrm{max} = 0.52$.
This angle is small so the flagella readily form bundles. The flexibilities are $\Fl = 2.5$ and $\Flh = 0.1$ so that uniflagellar swimming is stable. Snapshots in Fig.~\ref{fig:multiflag}(b) show the  tri- and quadriflagellar cases: flagella form rather tight bundles and the ultimate trajectories are straight. The open squares and circles on Fig. \ref{fig:multiflag}(a) show the translational and rotational speeds as a function of the number of flagella. Here the swimming speed increases weakly with $N$, but unlike in the peritrichous case has not saturated by $N=4$. However, the body rotation rate does increase almost linearly with the number of flagella. The asterisks show the translational speed results with hydrodynamic interactions turned off -- now the speed saturates by $N=4$. We do not fully understand the difference between the results with and without HI. Nevertheless, these results are generally consistent with the qualitative experimental observations of Darnton \emph{et al.} \cite{Darnton:2007ct}. 
We turn now to a simple model that sheds light on the observation that the swimming speed is not proportional to the number of flagella.

\color{black}

Consider a highly idealized model of a swimmer such that all motions and forces are restricted to one dimension.  The swimmer has $N$ independent flagella, each exerting a force $-F_p$ on the fluid, which in turn exerts a force $+F_p$ on the body. 
To write the force balance on the entire swimmer, we neglect hydrodynamic interactions and assume that the swimmer body and flagella move with constant speed $v_b^N$ and balance the total propulsion with the total drag (from body and $N$ flagella):
\begin{equation}
-\zeta_b v_b^N -N\zf v_b^N + N F_p = 0,
\end{equation}
where $\zeta_b$ is the friction coefficient for the body and $\zf$ is the translational friction coefficient for a flagellum moving along its axis. Solving for $v_b^N$ and normalizing by the uniflagellar value $v_b^1$  yields an expression for relative speed:
%
%
\begin{equation}
\frac{v_b^N}{v_b^1} =  \frac{N\left(\zeta_b + \zeta_f^\parallel\right)}{\zeta_b + N \zf}.
\label{eq:speed_multiplier}
\end{equation}
Estimating $\zeta_f^\parallel$ for our standard flagellum from RFT, the prediction from Eq. \ref{eq:speed_multiplier} shown by the blue line in Fig. \ref{fig:multiflag}(a) matches the simulation data reasonable well -- indeed for the polar case without HI it should (and does) yield nearly quantitative agreement. As $N\rightarrow\infty$, Eq.~\ref{eq:speed_multiplier} asymptotes to a constant value
\begin{equation}
\frac{v_b^N}{v_b^1} \rightarrow \frac{\zeta_b + \zeta_f^\parallel}{\zf}=1+\frac{\zeta_b}{\zf}.
\label{eq:speed_multiplier_infty}
\end{equation}
Thus
 there are diminishing returns on speed with more flagella -- although each flagellum provides more thrust, it also yields more drag. (Nevertheless, if the $\zeta_b\gg\zf$, the advange from swimming with a large number of flagella can be substantial.) 
In Appendix \ref{app:multiflag}, we describe results for more detailed model that accounts for flagellar rotation and body counter-rotation in a real swimmer. The resulting expression for $v_b^N/v_b^1$ has the same $N$-dependence as Eq.~\ref{eq:speed_multiplier} and also predicts that the relative body rotation rate $\omega_b^N/\omega_b^1=N$. This result is in good agreement with the numerical results for the polar case, which are reported as red circles in Figure \ref{fig:multiflag}.


\section{Conclusions}

\revision{
We have developed a model swimmer with a rigid body and one or more elastic flagella and  characterized the swimming dynamics in the parameter space of hook and flagellar flexibilities as well as flagellar multiplicity. In the multiflagellar case, most results address the dynamics then the flagella are at the corners of a regular tetrahedron inscribed in the spherical body.} 

\revision{In the uniflagellar case, straight swimming is stable only for a small range of low flexibilities. Modest increases in $\Flh$ lead to buckling of the hook and/or flagellum, and combined with higher $\Fl$ produce widely wobbling trajectories that are much slower than the straight case. For the bi- and triflagellar cases, we see that swimming is possible at low $\Flh$ with no bundling and at high $\Flh$ with bundling. In particular, by introducing the possibility of bundling of flagella with very weak hooks, the shift from uniflagellar to biflagellar morphology confers substantial stability, dramatically expanding the regime over which straight swimming can occur.  A further expansion in the parameter regime for stable swimming with flagellar bundles occurs as the morphology  shifts from biflagellar to triflagellar. 
 For the quadriflagellar swimmer,  swimming can occur only at high $\Flh$, and thus we see that the stability regime has reversed from that of the uniflagellar swimmer.  These results highlight the contrasting roles of flexibility in the locomotion of uni- and multiflagellar bacteria. }
 
\revision{Furthermore, we find that when $\Fl_h$ is sufficiently large, stable bundling and swimming is robust with respect to flagellar placement. Finally, we present computational results indicatign that swimming speed increase sublinearly with the number of flagella. A simple hydrodynamic theory is in reasonable agreement with the simulations. }


This work offers insight into the biology and function of natural bacterial swimmers, and may contribute to the design of artificial swimming devices.

\appendix

\section{Hydrodynamic tensors} \label{App:AppendixA}
We use the regularized solution to the Stokes equations with a quasi-Gaussian force density following Ref. \cite{Hernandez-Ortiz2005}:
\begin{equation}
\phi_\xi(\mathbf{r}) =  \left( \frac{\xi}{\sqrt{\pi}} \right)^3 \exp{(-\xi^2 r^2)} \left( \frac{5}{2} - \xi^2r^2 \right),
\end{equation}
where $r = |\mathbf{r}|$ and $\xi^{-1}$ is the length scale of the regularization. The velocity at any point in the fluid due to the elastic forces is given by:
\begin{equation}
\mathbf{u}(\mathbf{x}) = \sum_{k=1}^N \mathbf{S}_\xi(\mathbf{x}, \, \mathbf{x}_k) \cdot \mathbf{f}_k.
\label{eq:v_stokes}
\end{equation}
Here $\mathbf{S}_\delta$ is the regularized Stokeslet tensor:
\begin{equation}
\mathbf{S}_\xi(\mathbf{x}, \, \mathbf{x}_k) = \frac{1}{8\pi\eta} \left[ H_\xi^1(r) \boldsymbol{\delta} + \frac{\mathbf{r}_k \mathbf{r}_k}{r^2} H_\xi^2(r) \right],
\label{eq:reg_Stokeslet}
\end{equation}
\begin{equation}
H_\xi^{1,2}(r_k) = \frac{\mathrm{erf} (\xi r_k)}{r_k} \pm \frac{2}{\sqrt{\pi}}\exp{(-\xi^2r_k^2)} ,
\end{equation}
with $\mathbf{r} = \mathbf{x} - \mathbf{x}_k$ and $r_k = |\mathbf{r}_k|$. Note that we recover the conventional Stokeslet when $\xi^{-1} \rightarrow 0$, defined below:
\begin{equation}
	\mathbf{S}(\mathbf{x}, \mathbf{x}_k) = \frac{1}{8\pi\eta} \left[ \frac{\boldsymbol{\delta}}{r_k} + \frac{\mathbf{r}_k \mathbf{r}_k}{r_k^3} \right]
\end{equation}
Note that $\mathbf{S}$ is the Stokeslet used in Eq. \ref{eq:flow_sphere} to get the body flow field. We also define the rotlet tensor $\mathbf{R}$:
\begin{equation}
	\mathbf{R}(\mathbf{x}, \mathbf{x}_k) = \frac{1}{r_k^3} \left[ \mathbf{r}_k \right]_\times
\end{equation}
where $[ \cdot ]_\times$ denotes the cross product operator written in matrix form.

\section{Steric interactions}
\label{app:sterics}
The methodology described in this Appendix closely follows Adhyapak et al. \cite{Adhyapak:2015jk}. Accounting for every possible interaction among all components (body and flagellar nodes), we write the total steric force on the body and each flagellar node as:
\begin{equation}
	\mathbf{F}_b = \sum_i \mathbf{f}_{b,i}
	\label{eq:steric0b}
\end{equation}
\begin{equation}
	\mathbf{F}_i^\ster = \mathbf{f}_{i,b} + \sum_{j \neq i} \mathbf{f}_{i,j}
	\label{eq:steric0i}
\end{equation}
where the general notation $\mathbf{f}_{p,q}$ denotes the steric force \textit{on} component $p$ \textit{due to} contact with $q$. If we let $\mathbf{r}_s$ be the vector connecting closest points of contact (magnitude $r_s$ and direction $\hat{r}_s$), we obtain the general force equation from the potential energy (Eq. \ref{eq:U_LJ}):
\begin{equation}
	\mathbf{f}_{p,q}^s = -\frac{\mathrm{d}U_{LJ}(r_s)}{\mathrm{d}\mathbf{r}_s}.
\end{equation}
We know describe calculation of $\mathbf{r}_s$ for all interactions.

First we resolve body-flagellum interactions. For the $i$th flagellar edge, we find the point $\mathbf{x}_{i,b}^*$ on that edge closest to the body center $\mathbf{x}_b$ using a projection:
\begin{equation}
	\mathbf{x}_{i,b}^* = \mathbf{x}_{i-1} +  \mathcal{F} \left[ - (\mathbf{x}_{i-1} - \mathbf{x}_b) \cdot \mathbf{e}_i^3   \right] \mathbf{e}_i^3,
\end{equation}
\begin{equation}
	\mathcal{F}[h] = \left\{ \begin{array}{cl} 0, & \;\; h<0 \\ h, & \;\; h \in [0,l_{i,\eq}] \\ l_{i,\eq}, & \;\; h > l_{i,\eq}  \end{array}  \right. .
	\label{eq:filter}
\end{equation}
Here the truncation function $\mathcal{F}$ ensures that $\mathbf{x}_{i,b}^*$ is on edge $i$. If $r_s = |\mathbf{x}_{i,b}^* - \mathbf{x}_b| < \sigma$, we apply equal and opposite repulsive forces $\mathbf{f}_{i,b}^s = -\mathbf{f}_{b,i}^s$ on the body and edge $i$. On the former, we simply write $\mathbf{f}_{b,i} = -\mathbf{f}_{i,b}^s$. For the latter, we track translation only on adjacent nodes, so we decompose the force $\mathbf{F}^s$ acting on $\mathbf{x}_{i,b}^*$ to equivalent forces acting on $\mathbf{x}_{i-1}$ and $\mathbf{x}_i$:
\begin{equation}
	\mathbf{f}_{i-1,b} = \frac{h_{i,b}}{l} \mathbf{f}_{i,b}^s, \; \;\mathbf{f}_{i,b} = \frac{l-h_{i,b}}{l} \mathbf{f}_{i,b}^s,
	\label{eq:ster1}
\end{equation}
where $h_{i,b} = |\mathbf{x}_{i,b}^* - \mathbf{x}_i|$.

For the same flagellar edge $i$, we must also calculate the closest contact with \textbf{every} other flagellar edge $j$. We use Eqs. A1-A3 in Adhyapak et al. \cite{Adhyapak:2015jk} along with Eq. \ref{eq:filter}. We define $\mathbf{x}_{i,j}^*$ and $\mathbf{x}_{j,i}^*$ as the closest contacts located on edges $i$ and $j$ respectively. If $r_s = |\mathbf{x}_{i,j}^* - \mathbf{x}_{j,i}^*| < \sigma$ we again apply an equal and opposite steric force $\mathbf{f}_{i,j}^s = -\mathbf{f}_{j,i}^s$ to each edge, and decompose these forces into equivalent forces on adjacent nodes, as we did in Eq. \ref{eq:ster1}:
\begin{equation}
	\mathbf{f}_{i-1,j} = \frac{h_{i,j}}{l} \mathbf{f}_{i,j}^s, \; \;\mathbf{f}_{i,j} = \frac{l-h_{i,j}}{l} \mathbf{f}_{i,j}^s,
	\label{eq:ster2}
\end{equation}
\begin{equation}
	\mathbf{f}_{j-1,i} = -\frac{h_{j,i}}{l} \mathbf{f}_{i,j}^s, \; \;\mathbf{f}_{j,i} = -\frac{l-h_{j,i}}{l} \mathbf{f}_{i,j}^s,
	\label{eq:ster3}
\end{equation}
where $h_{i,j} = |\mathbf{x}_{i,j}^* - \mathbf{x}_i|$ and $h_{j,i} = |\mathbf{x}_{j,i} - \mathbf{x}_j|$.


\section{Solution algorithm and equations}
\label{app:numerics}

\revision{Time integration consists of two main parts: in the current state of the swimmer, we first use a projection algorithm to move the rigid body and flagellar nodes. After this constrained motion is applied, we then evaluate the torque balance on the edges and update  the local triads.}

\revision{
Let $\mathbf{X}$ contain all of the flagellar position vectors $\mathbf{x}_i^j$ and let $\mathbf{y}$ contain the position and orientation of the body and the positions of the nodes: %
\begin{equation}
	\mathbf{y} = \left[ \begin{array}{c} \mathbf{q}_b \\ \mathbf{x}_b \\ \mathbf{X} \end{array} \right].\label{eq:yvec}
\end{equation}
Following \cite{Bergou2008}, we write the rate of change of $\mathbf{y}$, $\dot{\mathbf{y}}$, as the sum of a unconstrained part (superscript $\dagger$) and a constrained part:}
\revision{
\begin{equation}
	\dot{\mathbf{y}} = \dot{\mathbf{y}}^\dagger + \mathbf{M} \cdot \left[ \nabla \mathbf{C}^\mathrm{T} \cdot \boldsymbol{\Lambda} \right]
	\label{eq:dae1}
\end{equation}
\begin{equation}
	\mathbf{C} = \mathbf{0}
	\label{eq:dae2}
\end{equation}
}
\revision{For the current flagellar configuration, we determine the velocity $\dot{\mathbf{y}}^\dagger$ from Eqs. \ref{eq:force_drag_flag}, \ref{eq:force_drag_body} and \ref{eq:quaternion}, and for flagellar anchors, we write $\mathbf{v}_0 = \mathbf{v}_b^\dagger +  \boldsymbol{\omega}_b^\dagger \times R_b\mathbf{e}_0^3$. The vector $\mathbf{C} \in \mathbb{R}^{N(M+2) + 1}$ comes from the constraint equations Eqs. \ref{eq:const1}, \ref{eq:const2} and \ref{eq:const3}. The quantity $\boldsymbol{\Lambda} \in \mathbb{R}^{N(M+2) + 1}$ is a vector of Lagrange multipliers (which yield constraint forces), and the gradient is taken with respect to $\mathbf{y}$. 
The quantity $\mathbf{M}$
\revision{$\in \mathbb{R}^{(3MN+7) \times (3MN+7)}$}
is a mobility matrix characterizing the hydrodynamic interactions between all swimming components:
\begin{equation}
	\mathbf{M} = \left[ \begin{array}{cc|c}
	\mathbf{M}_b^r & \mathbf{0} & \mathbf{M}_{b,i}^r \\
	\mathbf{0} & \mathbf{M}_b & \mathbf{M}_{b,i} \\
	\hline
	\mathbf{M}_{i,b}^r & \mathbf{M}_{i,b} & \mathbf{M}_{i}
	\end{array} \right]
\end{equation}
Here $\mathbf{M}_b^r$ and $\mathbf{M}_b$ are the rotational and translational mobilities of the body. Block matrices $\mathbf{M}_{b,i}^r$ and $\mathbf{M}_{b,i}$ capture the effect of flagellar motion on body, and conversely $\mathbf{M}_{i,b}^r$ and $\mathbf{M}_{i,b}$ capture the effect of body motion on flagella. The latter four matrices are calculated using Eqs. \ref{eq:flow_sphere} - \ref{eq:flagellar_flow_full}. The block matrix $\mathbf{M}_i$ captures all flagellum-flagellum interactions:
\begin{equation}
	\left[\mathbf{M}_i \right]_{j,k} = \left\{ \begin{array}{cl} 
	\boldsymbol{\zeta}_j^{-1}, &  \;\; j=k \\
	\mathbf{S}_\xi(\mathbf{x}_j,\mathbf{x}_k), & \;\; j \neq k
	\end{array}\right.
\end{equation}
where indices $j$ and $k$ run over all flagellar nodes. 
} 

\revision{Eqs. \ref{eq:dae1} and \ref{eq:dae2} form a $N(4M+2)+1$ system of differential-algebraic equations for the $N(4M+2)+1$ variables $[\mathbf{y}, \; \boldsymbol{\Lambda}]^\mathrm{T}$. 
The first step in time integration of these equations is to update $\dot{\mathbf{y}}^\dagger$:
}
%
\begin{equation}
	\mathbf{y}^\dagger(t_{n+1}) = \mathbf{y}(t_n) +  \dot{\mathbf{y}}^\dagger \Delta t
\end{equation}
%
Now, following \cite{Goldenthal2007}, we construct the following system, with constraints evaluated at $t_{n+1}$:
\begin{equation}
	\Delta t \left[ \nabla\mathbf{C} \cdot \mathbf{M} \cdot \nabla\mathbf{C}^\mathrm{T} \right]\boldsymbol{\Lambda} = \mathbf{C}.
	\label{eq:proj1}
\end{equation}
\begin{equation}
	\delta \mathbf{y} = - \Delta t \, \mathbf{M} \cdot \nabla \mathbf{C}^\mathrm{T} \cdot \boldsymbol{\Lambda}.
	\label{eq:proj2}
\end{equation}
The quantities $\boldsymbol{\Lambda}$ and $\delta \mathbf{y}$ are then determined iteratively as follows:
\begin{enumerate}
	\setlength{\itemsep}{-1em}
	\item Start with unconstrained soln: $\mathbf{y}^* = \mathbf{y}^\dagger(t_{n+1})$.
	\item While $|\mathbf{C}(\mathbf{y}^*)| > \epsilon$:
	\begin{enumerate}
			\item Solve Eq. \ref{eq:proj1} for $\boldsymbol{\Lambda}$.
	\item Solve Eq. \ref{eq:proj2} for $\delta \mathbf{y}$.
	\item Update: $\mathbf{y}^* + \delta \mathbf{y} \mapsto 
\mathbf{y}^* $.
	\end{enumerate}
	\item The solution satisfying the constraints is $\mathbf{y}(t_{n+1}) = \mathbf{y}^*$.
\end{enumerate}
The tolerance for satisfaction of the constraints is $\epsilon = 10^{-12}$.

\revision{Now the torque balance on the edges and kinematic equations are used to update the triads $\{\mathbf{e}_i^1,\mathbf{e}_i^2,\mathbf{e}_i^3\}$ for each flagellum. The angular velocity of edge $i$ is given by $\boldsymbol{\omega}_i$.  The torque balance Eq.~\ref{eq:torque_balance_edge} can be written 
\begin{equation}
	\boldsymbol{\omega_i} \cdot \mathbf{e}_i^3 = \zeta^{r-1} \left[ T_i^\el + \delta_{i1} (\mathbf{T}^\mot \cdot \mathbf{e}_i^3) \right], \label{eq:twisttorque}
\end{equation}
thus yielding one component of $\boldsymbol{\omega}_i$. Defining nodal velocities
\begin{equation}
 \mathbf{v}_i = [\mathbf{x}_i(t_{n+1}) - \mathbf{x}_i(t_n)] / \Delta t, 
\end{equation}
the other two components can be determined from the kinematic relations
	\begin{equation}
	\boldsymbol{\omega}_i \cdot \mathbf{e}_i^1 = -|\mathbf{x}_i - \mathbf{x}_{i-1}|^{-1} \; \mathbf{e}_i^2 \cdot (\mathbf{v}_i - \mathbf{v}_{i-1}),\label{eq:k1}
	\end{equation}
	\begin{equation}
	\boldsymbol{\omega}_i\cdot \mathbf{e}_i^2 = +|\mathbf{x}_i - \mathbf{x}_{i-1}|^{-1} \; \mathbf{e}_i^1 \cdot (\mathbf{v}_i - \mathbf{v}_{i-1}).\label{eq:k2}
	\end{equation}
Finally, the edges can be updated by integrating
\begin{equation}
\dot{\mathbf{e}}_i^\beta= \boldsymbol{\omega}_i \times \mathbf{e}_i^\beta, \;\; \beta = 1,2,3.\label{eq:updatees}
\end{equation}
The explicit Euler scheme is used to evolve these equations, followed by renormalization to unit length.} 
%

\section{Comparisons against resistive force theory}
\label{app:RFT}

\begin{figure} 
	\begin{minipage}{0.42\linewidth}
		\centering{\includegraphics[width=\linewidth]{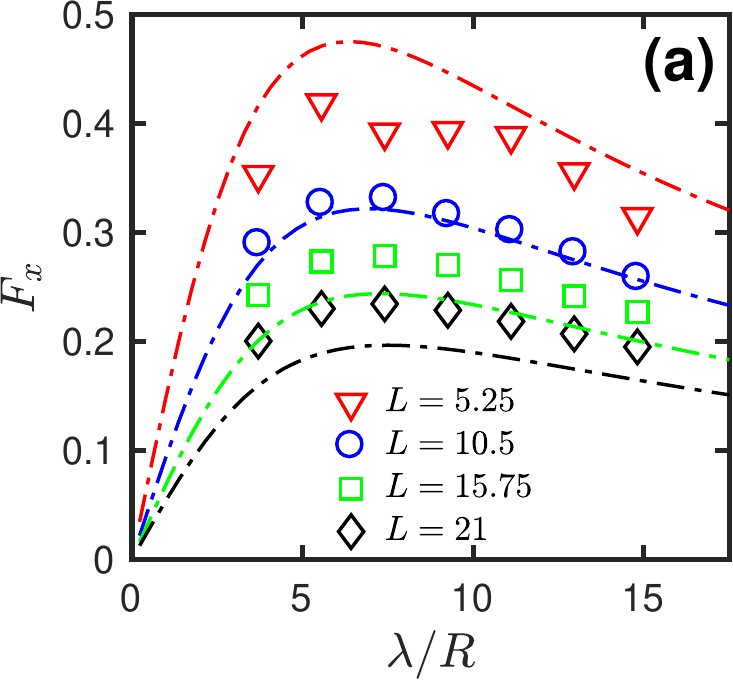}}
	\end{minipage}
	\hspace{5mm}
	\begin{minipage}{0.42\linewidth}
		\centering{\includegraphics[width=\linewidth]{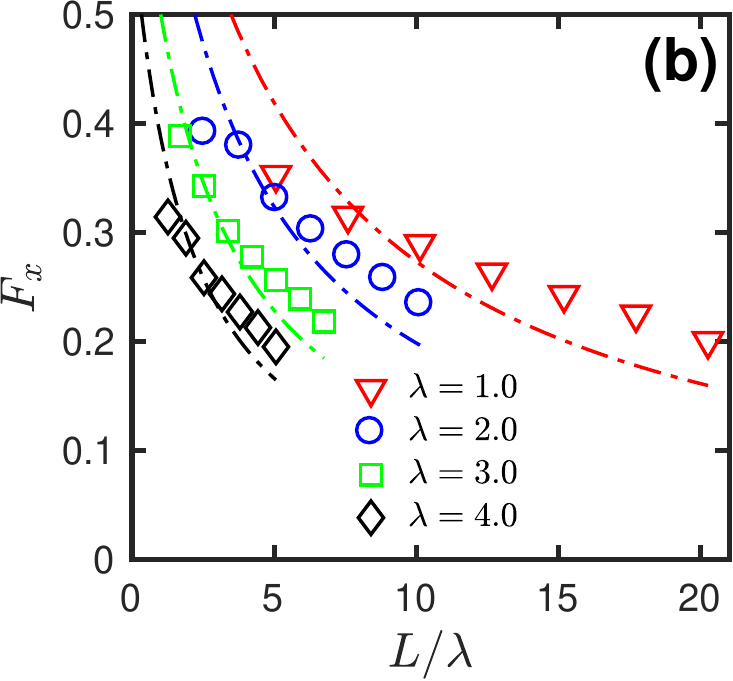}}
	\end{minipage}
	\caption{Propulsive force on body vs. flagellar helical geometry. The symbols are numerical simulations of the full elastic model with low flexibilities. The dashed lines are the RFT predictions using equations from Ref. \cite{Nguyen:2017jt}.}
	\label{fig:RFT}
\end{figure}

To validate our present formulation, we compare low-flexibility uniflagellar simulations using various flagellar geometries to our previous analytical predictions for swimmers with a rigid flagellum  \cite{Nguyen:2017jt} using resistive force theory. The propulsive forces on the body along the swimming direction, $F_x$, are summarized in Fig. \ref{fig:RFT}. Our current model results generally follow the same qualitative trend as the RFT calculations, with particularly good quantitative agreement for a few geometries. The moderate quantitative differences arising in all other cases is due to the inclusion of hydrodynamic interactions and elasticity in the current model. For the shorter flagella we test, end effects seem to play a large role, as evidenced by red triangles in Fig. \ref{fig:RFT}(a). In Fig. \ref{fig:RFT}(a), $F_x \rightarrow 0$ when $\lambda/R \rightarrow 0$ or $\lambda/R \rightarrow \infty$ as we expect because the flagellum becomes a ring or line, with neither capable of overcoming kinematic reversibility to generate thrust. In Fig. \ref{fig:RFT}(b). $F_x \rightarrow \infty$ as $L/\lambda \rightarrow 0$ because the drag vanishes faster than thrust. On the opposite end, $F_x \rightarrow 0$ as $L/\lambda \rightarrow \infty$ because there is too much drag on the flagellum. 

\section{Calculation of speed vs. number of flagella, including torque-thrust coupling}
\label{app:multiflag}


Here we generalize the analysis of Section \ref{sec:speed}. We again consider a 1D example where $N$ independent flagella all push the body in the same direction at the same anchor point. To generate this propulsion, a torque $T$ is applied by the body on each of the flagella to rotate them around their helical axes and generate thrust, with the flagella exerting an equal and opposite torque on the body, as well as a propulsive force $F_p$. Again assuming that the entire swimmer translates with constant velocity $v_b^N$, we write (in mobility formulation) the resulting force and torque balances on body and flagella as the sum of external dynamics and hydrodynamic drag:
\begin{equation}
	\left( \begin{array}{c} N F_p \\ -NT \end{array} \right)  - \left( \begin{array}{cc} \zeta_b & 0 \\ 0 & \zeta_{b,r} \end{array} \right) \left( \begin{array}{c} v_b^N \\ \omega_b^N \end{array} \right) = \left( \begin{array}{c} 0 \\ 0 \end{array} \right),
	\label{eq:toybodybal}
\end{equation}
\begin{equation}
	\left( \begin{array}{c} -F_p \\ T \end{array} \right)  - \left( \begin{array}{cc} \zf & \psi_f \\ \psi_f & \zfr \end{array} \right) \left( \begin{array}{c} v_b^N \\ \omega_f^N \end{array} \right) = \left( \begin{array}{c} 0 \\ 0 \end{array} \right)
	\label{eq:toyflagbal}.
\end{equation}
Here $\omega_b^N$ is the body rotation speed, $\omega_f^N$ the axial rotational speed of each flagellum, $\zfr$ the axial rotational friction coefficient, and $\psi_f$ the helical force-angular velocity coupling. From Eq. \ref{eq:toybodybal} we can immediately write:
\begin{equation}
	\omega_b^N = -\zeta_{b,r}^{-1}NT \; \; \rightarrow \; \; \frac{\omega_b^N}{\omega_b^1} = N.
\end{equation}
We then solve the remaining linear system of equations for $v_b^N$, $\omega_f^N$, and $F_p$. The solutions for the velocities are:
\begin{equation}
	v_b^N = -\frac{(\psi_f/\zfr)NT}{\zeta_b + N\left[\zf - \psi_f^2 / \zfr \right] },
\end{equation}
\begin{equation}
	\omega_f^N = \frac{T}{\zfr} \left[1 + \frac{N \psi_f^2 / \zfr}{\zeta_b + N\left[\zf - \psi_f^2 / \zfr \right] } \right].
\end{equation}
The body translation and flagellar rotation relative to the uniflagellar values $v_b^1$ and $\omega_b^1$ are then:
\begin{equation}
	\frac{v_b^N}{v_b^1} = N \frac{\zeta_b + \zf - \psi_f^2/ \zfr}{\zeta_b + N\left[\zf - \psi_f^2 / \zfr \right] } < N,
	\label{eq:toy_relv}
\end{equation}
\begin{equation}
	\frac{\omega_f^N}{\omega_f^1} =\left[ 1 - \frac{\psi_f^2 / \zfr}{ \zeta_b + \zf } \right]  \left[ 1 + \frac{N \psi_f^2 / \zfr}{\zeta_b + N\left[\zf - \psi_f^2 / \zfr \right] } \right].
	\label{eq:toy_relw}
\end{equation}
Because the flagellar mobility matrix is positive definite, $\zf \zfr - \psi_f^2 >0$ and the inequality in Eq. \ref{eq:toy_relv} always holds. We note that as $N \rightarrow \infty$, $v_b^N/v_b^1$ and $\omega_f^N/\omega_f^1$ become constant; in particular we see the same diminishing return on body speed as in Eq. \ref{eq:speed_multiplier}, although $\omega_b^N \rightarrow \infty$.

\section*{Acknowledgments}
This material is based on work supported by the National Science Foundation under grant no. PHY-1304942 and by the Graduate Engineering Research Scholars program of the University of Wisconsin-Madison.


%

\end{document}